\RequirePackage{lineno}
\documentclass[ALICE,manyauthors]{cernphprep}
\usepackage[comma,square,numbers,sort&compress]{natbib}
\usepackage{graphicx}
\usepackage{dcolumn}
\usepackage{bm}
\usepackage{color}
\usepackage{hyperref}
\usepackage{multirow}
\usepackage{multicol}
\usepackage{booktabs}
\usepackage{epstopdf}
\usepackage{rotating}
\usepackage{tabularx}
\usepackage{epstopdf}
\usepackage{afterpage}

\newcolumntype{Y}{>{\centering\arraybackslash}X}

\newcommand{\Om}{$\Omega^-$}
\newcommand{\Mo}{$\overline{\Omega}^+$}
\newcommand{\X}{$\Xi^-$}
\newcommand{\Ix}{$\overline{\Xi}^+$}
\newcommand{\Xis}{$\Xi^{\pm}$}
\newcommand{\Oms}{$\Omega^{\pm}$}
\newcommand{\meanpt}{$\langle p_\mathrm{T}\rangle$}

\newcommand{\seven}{$\sqrt{s}~=~7$~TeV}

\newcommand{\fivenn}{$\sqrt{s_{\rm{NN}}}~=~5.02$~TeV}
\newcommand{\twosevensixnn}{$\sqrt{s_{\rm{NN}}}~=~2.76$~TeV}

\newcommand{\GeVc}{GeV/$c$}

\newcommand{\GeVmass}{GeV/$c^2$}
\newcommand{\MeVmass}{MeV/$c^2$}

\newcommand{\pip}          {\ensuremath{\pi^{+}}}
\newcommand{\pim}          {\ensuremath{\pi^{-}}}
\newcommand{\pipm}          {\ensuremath{\pi^{\pm}}}
\newcommand{\kap}          {\ensuremath{\mathrm{K}^{+}}}
\newcommand{\kam}          {\ensuremath{\mathrm{K}^{-}}}
\newcommand{\kapm}          {\ensuremath{\mathrm{K}^{\pm}}}
\newcommand{\p}               {$\rm p$}
\newcommand{\pbar}         {$\rm\overline{p}$}
\newcommand{\kzero}        {\ensuremath{{\rm K}^{0}_{\rm S}}}
\newcommand{\kstar}        {\ensuremath{{\rm K}^{*0}}}
\newcommand{\lmb}          {\ensuremath{\Lambda}}
\newcommand{\almb}         {\ensuremath{\overline{\Lambda}}}

\newcommand{\pp}           {pp\xspace} 

\newcommand{\PbPb}         {\mbox{Pb--Pb}}
\newcommand{\pPb}          {\mbox{p--Pb}}

\newcommand{\dNdeta}       {\ensuremath{\mathrm{d}N_\mathrm{ch}/\mathrm{d}\eta}}

\newcommand{\avdNdeta}       {\ensuremath{\left<\mathrm{d}N_\mathrm{ch}/\mathrm{d}\eta\right>}}

\newcommand{\dNdy}         {\ensuremath{\mathrm{d}N/\mathrm{d}y} }

\newcommand{\s}            {\ensuremath{\sqrt{s}}}
\newcommand{\pt}           {\ensuremath{p_\mathrm T\,}\xspace}

\newcommand{\ppi}          {\ensuremath{{\rm p}/\pi}}
\newcommand{\kpi}          {\ensuremath{{\rm K}/\pi}}

\newcommand{\mt}           {\ensuremath{m_{\rm T}}}

\newcommand{\avpT}         {\ensuremath{\left< \pt \right>}\xspace}

\newcommand{\gevc}         {\ensuremath{{\rm GeV}/c}}

\newcommand {\dEdx}      {d\textit{E}/d\textit{x}\xspace}

\newcommand {\proton}     		{\ensuremath{p}}

\newcommand {\pion}    	  		{\ensuremath{\pi}}
\newcommand {\kaon}       		{\ensuremath{K}}
\newcommand {\KTopi}      		{\kaon/\pion}
\newcommand {\pTopi}      		{\proton/\pion}

\newcommand{\pPiplus}{\ensuremath{{\pi}^{+}}\xspace}
\newcommand{\pPiminus}{\ensuremath{{\pi}^{-}}\xspace}

\newcommand{\pKplus}{\ensuremath{{\rm K}^{+}}\xspace}
\newcommand{\pKminus}{\ensuremath{{\rm K}^{-}}\xspace}

\newcommand{\pProton}{\ensuremath{\rm p}\xspace}
\newcommand{\apProton}{\ensuremath{\overline{\rm p}}\xspace}

\newcommand{\pLambda}{\ensuremath{\Lambda}\xspace}

\newcommand{\LtoKzero}{\ensuremath{\Lambda}/\ensuremath{{\rm K}^{0}_{S}}\xspace}
\newcommand{\apLambda}{\ensuremath{\overline{\Lambda}}\xspace}
\newcommand{\sXi}{\ensuremath{\Xi}\xspace}
\newcommand{\pXi}{\ensuremath{\Xi^{-}}\xspace}
\newcommand{\apXi}{\ensuremath{\overline{\Xi}^{+}}\xspace}
\newcommand{\sOmega}{\ensuremath{\Omega}\xspace}
\newcommand{\pOmega}{\ensuremath{\Omega^{-}}\xspace}
\newcommand{\apOmega}{\ensuremath{\overline{\Omega}^{+}}\xspace}

\newcommand{\betaT}{\ensuremath{\langle \beta_{T}\rangle}\xspace}
\newcommand{\Tkin}{\ensuremath{T_{kin}}\xspace}

\begin{document}

\begin{titlepage}
\PHyear{2018}
\PHnumber{209}      
 \PHdate{23 July}  

\title{Multiplicity dependence of light-flavor hadron production \\ in pp collisions at \s\ = 7 TeV}
\ShortTitle{Multiplicity dependence of identified hadron production in pp}
\Collaboration{ALICE Collaboration\thanks{See Appendix~\ref{app:collab} for the list of collaboration members}}
\ShortAuthor{ALICE Collaboration} 

\begin{abstract} 

Comprehensive results on the production of unidentified charged particles, \pipm, \kapm, \kzero, K*(892)$^{0}$, \p, \pbar, $\phi$(1020), $\lmb$, $\almb$, \X, \Ix, \Om\ and \Mo hadrons in proton-proton (pp) collisions at \s\ = 7 TeV at midrapidity ($|y| < 0.5$) as a function of charged-particle multiplicity density are presented. In order to avoid auto-correlation biases, the actual transverse momentum (\pt) spectra of the particles under study and the event activity are measured in different rapidity windows. In the highest multiplicity class, the charged-particle density reaches about 3.5 times the value measured in inelastic collisions. While the yield of protons normalized to pions remains approximately constant as a function of multiplicity, the corresponding ratios of strange hadrons to pions show a significant enhancement that increases with increasing strangeness content. Furthermore, all identified particle to pion ratios are shown to depend solely on charged-particle multiplicity density, regardless of system type and collision energy. 
The evolution of the spectral shapes with multiplicity and hadron mass shows patterns that are similar to those observed in p-Pb and Pb--Pb collisions at LHC energies. The obtained \pt\ distributions and yields are compared to expectations from QCD-based 
pp event generators as well as to predictions from thermal and hydrodynamic models. These comparisons indicate that 
traces of a collective, equilibrated system are already present in high-multiplicity pp collisions. 

\end{abstract}
%
%
\end{titlepage}
\setcounter{page}{2}
%


\section{\label{secIntroduction}Introduction}

Recently, several collective phenomena have been observed in high-multiplicity \pp\ and p-Pb collisions that are reminiscent of observations attributed to the creation of a medium in thermal and kinematic equilibrium in Pb--Pb collisions. In p--Pb collisions, these include the observation of double-ridge structures on the near and away side in two-particle correlation studies~\cite{Abelev:2012ola}, non-vanishing $v_2$ coefficients in multi-particle cumulant studies \cite{Khachatryan:2015waa}, mass dependent hardening of identified particle \pt\ spectra~\cite{ABELEV:2013wsa,Abelev:2013haa,Abelev:2014uua} and consistency of integrated particle yield ratios with thermal model expectations at high multiplicities~\cite{Adam:2015vsf}. 

While double-ridge structures have already been observed in high-multiplicity pp collisions~\cite{Aad:2015gqa}, a comprehensive study of
identified-particle hadrochemistry as well as of the corresponding kinematics after hadronization has not yet been performed in these
collisions: such a study is the main topic covered in this paper. The investigation of mass dependent effects as expected within a hydrodynamic 
evolution scenario requires the measurement of several particle species such as the ones presented here and relies on the excellent particle identification capabilities provided by the ALICE detector.

While similarities in the production of light-flavor hadrons between p-Pb and Pb--Pb collisions at comparable event multiplicities have been discussed previously~\cite{Abelev:2013haa,Adam:2015vsf}, the measurements presented here allow a unique comparison of the observables with several QCD-inspired event generators such as PYTHIA~\cite{Skands:2014pea} and EPOS~\cite{Pierog:2013ria}. Traditionally, bulk particle production in heavy-ion collisions is described on the basis of equilibrium many-body theories such as hydro- and thermodynamics (see for instance \cite{deSouza:2015ena,Andronic:2017pug} and references therein). The continuous transition of light-flavor hadron measurements from pp to p-Pb and Pb--Pb collisions as a function of event multiplicity thus links the dynamic production of particles in individual 2$\rightarrow$2 QCD parton-parton scattering processes and subsequent hadronization as an underlying equilibration mechanism to a thermodynamic description of the system.

In a recent letter \cite{Adam:2016emw}, the ALICE Collaboration reported the multiplicity dependent enhancement of strange (\kzero, $\lmb$ and $\almb$) and multi-strange (\X, \Om, \Ix\ and \Mo) particle production in pp collisions at  \s\ = 7 TeV. In this paper, those results are complemented by the measurement of  \pipm, \kapm, \p, \pbar, K*(892)$^{0}$, and $\phi$(1020), as well as by an extended discussion on \pt-differential and \pt-integrated particle ratios and model comparisons. For the sake of brevity, in this work, ($\pi^{+}+\pi^{-}$) and ($\mathrm{K}^{+}+\mathrm{K}^{-}$) will be denoted by \pipm and \kapm, while \p\ refers to 
(\p+\pbar) unless otherwise stated. In addition, ($\Xi^{-}+\overline{\Xi}^{+}$) and ($\Omega^{-}+\overline{\Omega}^{+}$) will be denoted by $\Xi$ and $\Omega$, while \lmb\ refers solely to the particle and not the antiparticle unless otherwise stated. 
Finally, (K*(892)$^{0}$+$\overline{\rm{K}^{*}(892)}^{0}$) and $\phi$(1020) will be denoted simply by \kstar\ and $\phi$ throughout this document. 
The paper is organized as follows. In section~\ref{secAnalysis}, the details of the analysis techniques and of the event classification are described. The results are given in section~\ref{secResults} in which the transverse momentum spectra as well as the extraction of the \pt-integrated yields and average transverse momenta are presented. Detailed model comparisons and an interpretation of the results are presented and discussed in section~\ref{secDiscussion}.

\section{\label{secAnalysis}Analysis}

For this analysis, data collected by ALICE in the LHC pp run of the year 2010 are used. In total, the analysis is based on up to 281 million minimum-bias events, corresponding to an integrated luminosity of 4.5~nb$^{-1}$. 
A detailed description of the ALICE apparatus and of its performance can be found in \cite{Aamodt:2008zz,Abelev:2014ffa}. The main subdetectors used in this analysis are the Inner Tracking System (ITS)~\cite{its,Aamodt:2010aa}, the Time Projection Chamber (TPC)~\cite{Alme:2010ke}, the Time-Of-Flight detector (TOF)~\cite{Akindinov2013},  and the V0 scintillator hodoscopes~\cite{1748-0221-8-10-P10016}. 
All tracking detectors are positioned inside a magnetic field $B$~=~0.5~T. 

The innermost barrel detector is the ITS, consisting of six cylindrical layers of high-resolution silicon tracking detectors using three different technologies. The two innermost layers are based on silicon pixel technology (SPD) with digital readout. The four outer layers, made of drift (SDD) and strip (SSD) detectors provide analogue readout and thus allow for particle identification via specific energy loss. The ITS, used as a standalone tracker, enables the reconstruction and identification of low momentum particles down to 100~MeV/$c$ that do not reach the TPC.

The TPC is a large cylindrical drift detector of radial and longitudinal dimensions of approximately 85~cm~$< r <$~250~cm and  $-$250~cm~$< z <$~ 250~cm, respectively. As the main tracking device, it thus provides full azimuthal acceptance for tracks in the pseudorapidity region $|\eta| < 0.9$. In addition, it provides particle identification via the measurement of the specific energy loss d$E$/d$x$. At low transverse momenta (\pt\ $\lesssim$ 1.0~\gevc), the d$E$/d$x$ resolution of 5.2\% for a minimum ionizing particle allows a track-by-track identification, while at high transverse momenta (\pt\ $\gtrsim$ 2.0~\gevc) the overlapping energy losses can still be statistically distinguished using a multi-gaussian fit to the d$E$/d$x$ distributions.

Further outwards in radial direction from the beam-pipe and located at a radius of approximately 4~m, the TOF measures the time-of-flight of the particles, providing particle identification over a broad range at intermediate transverse momenta (0.5 $\lesssim$ \pt\ $\lesssim$ 2.7~\gevc). It is a large-area array of multigap resistive plate chambers with an intrinsic time resolution of 50 ps. The total time resolution includes contributions from the start time determination and amounts to about 120~ps in pp collisions. As described in detail in~\cite{Adam:2016ilk}, the start time contribution to the total time resolution improves with increasing number of hits in the TOF in a given event, thus leading to a slight dependence on the event multiplicity and results in a total time resolution of about 100~ps for the highest multiplicities under study. 

The V0 detectors are two scintillator hodoscopes that are located on either side of the interaction region at $z = 3.3$~m and $z$~=~$-0.9$~m, respectively. They cover the pseudorapidity region 2.8~$< \eta <$~5.1 and $-$3.7~$< \eta <$~$-$1.7 in full azimuth and are employed for triggering, background suppression, and event-class determination.

Measurements of unidentified and identified primary particles are reported. Primary particles are defined as
any hadron with a mean proper lifetime that is of at least 1~cm/$c$ either produced directly in the interaction or emerging from decays of particles with lifetime shorter than 1~cm/$c$ and excluding particles from interactions with the detector material~\cite{ALICE-PUBLIC-2017-005}. The criteria for the selection of primary tracks for \pipm, \kapm, \p\ and \pbar\ as well as for the decay products of K*$^{0}$ and $\phi$ follow the  procedures described in~\cite{Adam:2015qaa}.
All measurements  are corrected for detector acceptance and  reconstruction efficiency using Monte Carlo events generated 
with PYTHIA6 Perugia 0 \cite{Sjostrand:2006za,Skands:2010ak} and propagated through the full ALICE geometry with 
GEANT3 \cite{Geant:1994zzo}. These events are then reconstructed using the same techniques employed in the case of real data. 
The corresponding detector acceptance and reconstruction efficiencies are found to be multiplicity-independent within 1\%
and thus the multiplicity-integrated values are used for all event classes to minimize statistical fluctuations.

\subsection{\label{subsecEventSelection}Event selection and classification}

The data were collected using a minimum-bias trigger requiring a hit either in the A or C side of the V0 (denoted in what follows as V0A or V0C, respectively) or in the SPD, in coincidence with the arrival of proton bunches from both directions. Contamination from beam-gas events is removed offline by using timing information from the V0, which has a time resolution better than 1~ns. Events in which pile-up or beam-gas interactions occurred are also rejected by exploiting the correlation between the number of pixel hits and the number of SPD tracklets. Interactions used for the data analysis are further required to have a reconstructed primary vertex within $|z| < 10$~cm, where $z$ is in the direction of the beam. Events containing more than one reconstructed vertex are tagged as pile-up occurring within the same
 bunch-crossing and discarded for the analysis, with up to 10\% of all events being tagged in the highest-multiplicity event class considered for analysis. The pile-up tagging was estimated to be efficient enough so that the residual pile-up remaining in the 
analysed event sample is of no more than $10^{-4}$ to $10^{-2}$ for the lowest and highest multiplicity classes, respectively. The systematic uncertainty associated to pileup rejection was estimated to be smaller than 1\% and is therefore not a dominant source of uncertainty for any of the analyses reported here. 

The measurements shown here correspond to an event class (INEL $>$ 0) in which at least one charged particle is produced in the pseudorapidity interval $|\eta| < 1$ with respect to the beam, corresponding to about 75\% of the total inelastic cross-section. In order to study the multiplicity dependence of light-flavor hadron production, the sample is divided into event classes based on the total charge deposited in both of the V0 detectors (V0M amplitude). 
The V0M amplitude is found to be linearly proportional to the total number of charged particles produced in the pseudorapidity window corresponding to the acceptance of the V0 scintillators.


Table~\ref{tab:OverviewEventClasses} also lists the average charged-particle pseudorapidity densities $\langle {\rm d }N_{\rm ch}/{\rm d }\eta \rangle$ within $|\eta|<0.5$ for the different event multiplicity classes. The relative standard deviations of the corresponding distributions range from 68\% to 30\% of the average $\langle {\rm d }N_{\rm ch}/{\rm d }\eta \rangle$ for the event class with the lowest average multiplicity to the one with the highest, respectively. These are obtained based on the reconstruction of SPD tracklets which have an acceptance of  \pt~$\gtrsim 50$~MeV/$c$. The measurement has been fully corrected for acceptance, tracking and vertexing efficiency as well as for contamination from secondary particles and combinatorial background. 
 Further details can be found in~\cite{ALICE:2012xs,Adam:2015pza}. In addition, all quantities reported in this work are 
corrected for event detection efficiencies using a data-driven unfolding method. This correction is negligible for high-multiplicity event classes but is of up to $\sim$13\% in multiplicity class X. 
The resulting percentages of the total INEL$>$0 cross-section, $\sigma/\sigma_{\rm INEL>0}$, are also reported in Tab.~\ref{tab:OverviewEventClasses}.
 These values are reported after event detection 
efficiency corrections and do not match the integer boundaries that were used in analysis, e.g. high-multiplicity event classes such as I and II were 
selected as 0-1\% and 1-5\% for analysis but event losses at low-multiplicity compress these fractions into 0-0.95\% and 0.95-4.7\% of the true INEL $>$ 0 cross-section. 
The analysis-level selection percentiles have been omitted as they are detector-dependent quantities.

\begin{table}[htbp]
\centering
\begin{tabular}{llllll} \hline \hline
Multiplicity class & I               & II              & III             & IV              & V               \\ \hline
$\sigma/\sigma_{\rm INEL>0}$      & 0-0.95\%        & 0.95-4.7\%      & 4.7-9.5\%       & 9.5-14\%        & 14-19\%         \\
$\langle {\rm d }N_{\rm  ch}/{\rm d }\eta \rangle$       & $21.3 \pm 0.6$  & $16.5 \pm 0.5$  & $13.5 \pm 0.4$  & $11.5 \pm 0.3$  & $10.1 \pm 0.3$  \\ \hline \hline
Multiplicity class                       & VI              & VII             & VIII            & IX              & X               \\ \hline
$\sigma/\sigma_{\rm INEL>0}$      & 19-28\%         & 28-38\%         & 38-48\%         & 48-68\%         & 68-100\%        \\
$\langle {\rm d }N_{\rm ch}/{\rm d }\eta \rangle$       & $8.45 \pm 0.25$ & $6.72 \pm 0.21$ & $5.40 \pm 0.17$ & $3.90 \pm 0.14$ & $2.26 \pm 0.12$ \\ \hline \hline
\end{tabular}
  ~\newline
\caption[Event classes]{Event multiplicity classes used in the analysis, their corresponding fraction of the INEL$>$0 cross-section ($\sigma/\sigma_{\rm INEL>0}$) and their corresponding $\langle {\rm d }N_{\rm ch}/{\rm d }\eta \rangle$ in $|\eta|<0.5$. The value of $\langle {\rm d }N_{\rm ch}/{\rm d }\eta \rangle$ in the inclusive INEL$>$0 class is $5.96 \pm 0.23$. The uncertainties are the quadratic sum of statistical and systematic contributions. Table from~\cite{Adam:2016emw}.}
\label{tab:OverviewEventClasses}
\end{table}

\begin{figure*}[t!]
\center{\includegraphics[width=0.475\textwidth]{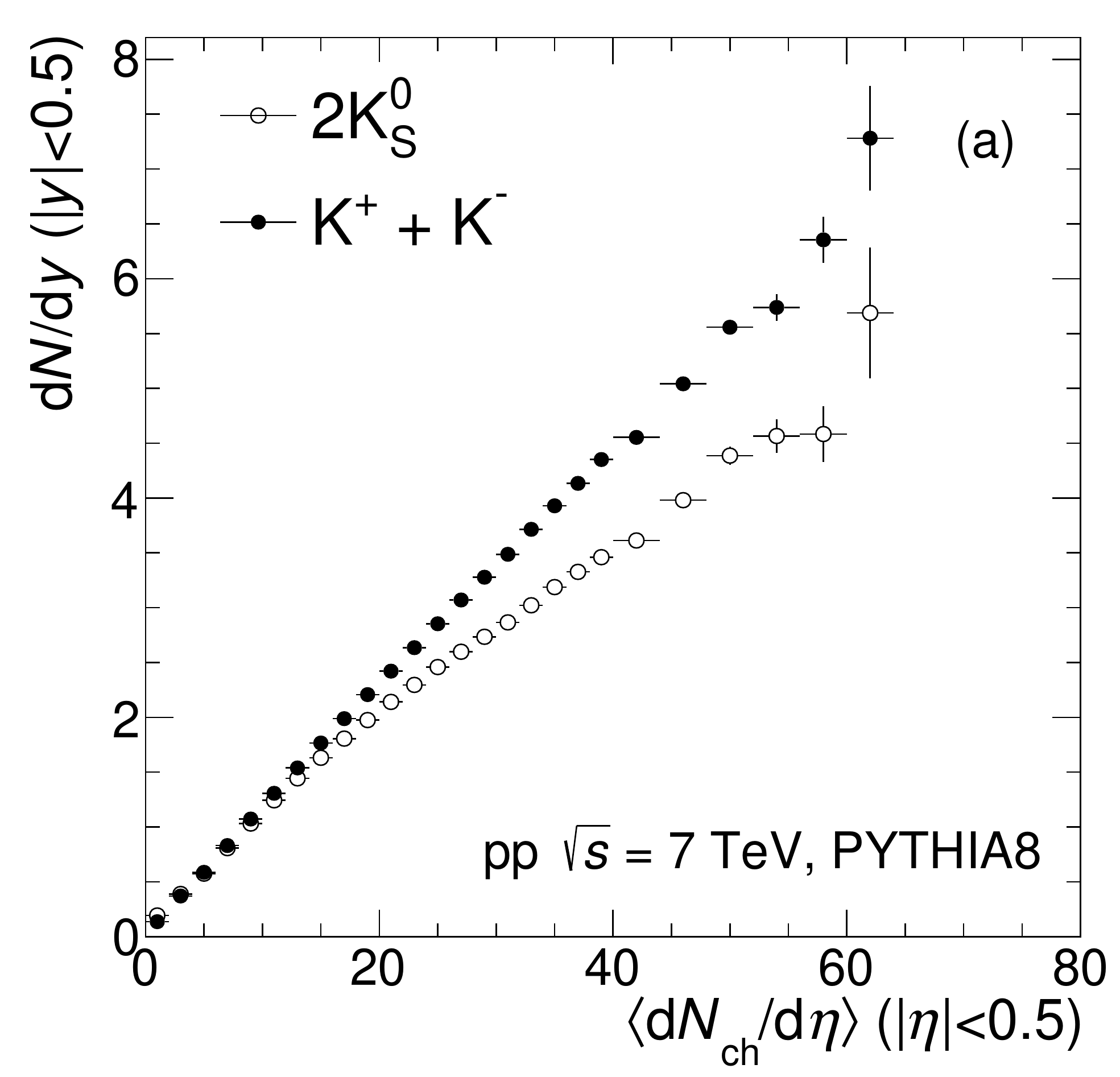}\includegraphics[width=0.475\textwidth]{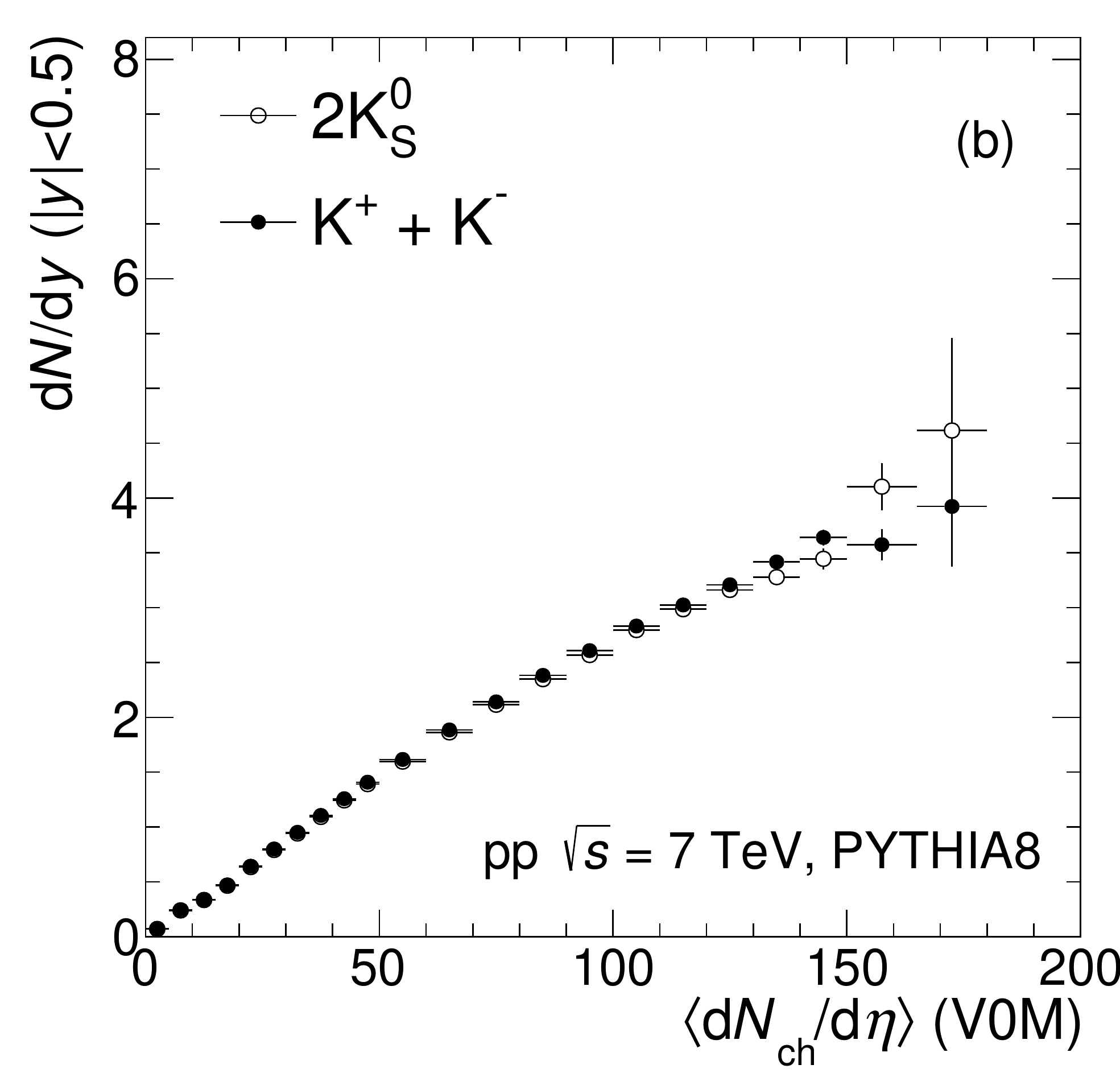}}
    \caption{Multiplicity dependence of charged and neutral kaon yields obtained using (a) mid-pseudorapidity charged particle multiplicities ($|\eta|<0.5$) and (b) the charged particle multiplicities within
    the pseudorapidity range corresponding to the V0A and V0C detectors (denoted by V0M, corresponding to $-3.7<\eta<-1.7$ and $2.8<\eta<5.1$) in PYTHIA8 simulations of inelastic pp 
    collisions at \seven.}
    \label{fig:selectionbiases}
\end{figure*}

In previous studies, event classification was based on  midrapidity charged-particle densities~\cite{Abelev2012165, Abelev2013371, Adam:2015pza},

as opposed to the forward and backward-pseudorapidity based selection utilized in this work. This choice is motivated by the fact that performing multiplicity selection and data 
analysis in the same pseudorapidity range may lead to auto-correlation biases and unphysical results. More specifically, hadrochemistry is significantly 
altered by selection biases, as exemplified by the progression of charged and neutral kaon abundances with multiplicity. If midrapidity-based selections 
were used, the integrated yields of \kapm\ for high-multiplicity events would be higher than the ones for \kzero\  because of the requirement of high charged-particle 
yields in the same pseudorapidity range. 
Conversely, if selection is performed with charged particle yields in a different pseudorapidity range than the one in which \kapm\ and \kzero\ production rates
are measured, similar amounts of charged and neutral kaons would be found across multiplicity, as expected due to
their similar masses. This 
can be readily tested in Monte Carlo simulations, as shown, for instance, in Fig.~\ref{fig:selectionbiases}, where the charged and neutral kaon yields in pp collisions simulated
with the PYTHIA8 event generator using the Monash 2013 tune~\cite{Sjostrand:2007gs, Skands:2014pea} are studied as a function of either midrapidity or forward/backward 
pseudorapidity charged-particle multiplicity. 
A significant bias towards charged kaons is observed in the former case, while the latter selection preserves the expected neutral-to-charged kaon ratio of approximately 
unity. 

This discrepancy for charged and neutral kaons is understood to be a consequence of performing selections on charged-particle multiplicities whose 
probability distributions exhibit a rapid decrease and have low 
average values. Under such circumstances, any multiplicity selections are likely to isolate fluctuations of charged-particle yields in the reference region of 
phase space rather than uniformly affecting all particle species regardless of their charge. In the particular case of \kapm\ and \kzero\ production rates, residual 
differences in these kaon yields still arise from resonance decay products, given that $\phi$ mesons decay preferentially into charged kaons. 
However, Monte Carlo studies show that these different feed-down contributions introduce differences in \kapm\ and \kzero\ yields of no more than 1-2\%, further corroborating 
the need to take into account the much larger selection bias effects shown in Fig.~\ref{fig:selectionbiases}. 

However, while multiplicity selections performed in different phase space regions will avoid selection biases, they are also naturally susceptible to the mid- to 
forward/backward pseudorapidity multiplicity correlation, which for small systems is not as strong as for nuclear collisions~\cite{PhysRevC.91.064905}. 
This has the 
consequence that the reach in midrapidity charged-particle densities is restricted in comparison to same-phase-space selections: when selecting high charged-particle
multiplicities in forward/backward pseudorapidity detectors, midrapidity $\langle {\rm d }N_{\rm ch}/{\rm d }\eta \rangle$ will eventually saturate, while it will still increase 
if event selection is performed with detectors at midrapidity. 
Furthermore, the V0 scintillators
that are used in this work for forward/backward charged-particle detection and event classification introduce an imperfect detector response into the analysis. In order
to minimize potential biases coming from these factors, 
all observables are studied as a function of charged-particle density at midrapidity $\langle$\dNdeta$\rangle$. By doing so, both \dNdeta\ and 
the variables under study are similarly folded with mid- to forward/backward pseudorapidity multiplicity correlations as well as the detector response within a given event class. 
This allows comparing results from this study with predictions from models by performing selections on charged-particle production in 
the acceptance of the V0 and it has been verified that any residual effect because of the finite detector resolution is negligible.

\subsection{\label{subsecunidentified}Unidentified Charged Particles}

Spectra of positively and negatively charged particles were obtained separately and summed afterwards.
The differences between the final spectra for particles and antiparticles were found to be around ~1.5\%.
The unidentified charged particles were reconstructed using the combined information from ITS and TPC. The \pt~range of the spectra in all multiplicity classes based on the V0M amplitudes is 0.16-40\ GeV/$c$ and the pseudorapidity was limited to $|\eta| < 0.5$. 
This pseudorapidity limit allows a comparison of the charged hadron spectrum with the sum of pions, kaons and protons analyzed in the rapidity range $|y| < 0.5$ by transforming them to the corresponding pseudorapidity window with the appropriate Jacobian ${{\rm d}^2{N} \over {\rm d}y{\rm d}\pt} = {E \over p} {{\rm d}^2{N} \over {\rm d}\eta{\rm d}\pt}$ for each \pt-interval. This cross-check showed a difference of less than 5\% which is consistent within the non-common systematics with the expected contributions from electrons, muons, and heavier charged baryons that are counted in addition to pions, kaons, and protons in the charged hadron spectrum.

The contribution from secondary particles was calculated in the same manner as described in detail in Section \ref{subsecPiKP}. The additional corrections based on FLUKA \cite{Ferrari:2005zk,BOHLEN2014211} for kaons and anti-protons, which are needed for those specific identified particle measurements in order to account for an imperfect description of absorption cross-section in the detector material, were found to have negligible impact on the unidentified charged particle spectra and were therefore not applied. 

The systematic uncertainties are summarized in Table \ref{tab:sysunidentified}. The multiplicity dependence of the tracking efficiency and the feed-down correction were found to be less than 1\% and were included in the final systematic uncertainty. The total systematic uncertainty is \pt~dependent, with values around 6-7\% up to 20 GeV/$c$. It reaches 9.6\% for the highest \pt~bin. The main contributions to the total systematic uncertainty come from the global tracking efficiency (5\%) and the parameter variation for the track selection criteria (3-7\%). The other sources have a \pt~dependent contribution of less than 2\% each. The systematic uncertainties related to the dependence of the reconstruction efficiency on the MC event generator and the particle composition have been studied as described in \cite{Abelev:2013ala}. All sources of uncertainty are assumed to be uncorrelated and the total uncertainty was calculated as the quadratic sum of the different contributions. The systematic uncertainty contribution that is uncorrelated across multiplicities was 
estimated to be 2.1\% for all the V0M multiplicity bins over the entire \pt~range.

\begin{table*}[ph!]
  \begin{tabularx}{\textwidth}{p{5.3cm}*{3}{Y}}
\hline\hline
    {\bf Source} &  \multicolumn{3}{c}{Uncertainty (\%)}   \\
\hline
    \pt\ (\GeVc)   &   0.16 & 3.0 & 40.0  \\
    \cmidrule(l{5pt}){2-4} 

    Correction for secondaries & 0.2  & 0.2 & 1 \\
    Particle composition in secondaries & 1.7 & 1.3 & 0.8 \\
    Material budget & 1.5 & negl. & negl.  \\
    Global tracking efficiency & 5 & 5 & 5  \\
    Particle composition & 1.7 & 2 & 2 \\
    Track selection & 4 & 2.8 & 7.4 \\
    MC event generator & 1.1 & 1.8 & 2 \\
    \pt\ resolution & negl. & negl. & 0.5 \\
    Efficiency multiplicity dependence & 1 & 1 & 1 \\
\hline

    Total & 7.2 & 6.6 & 9.6  \\ 
    
    Total ($N_{ch}$-independent) & 6.9 & 6.3 & 9.4 \\ 
\hline\hline
 \end{tabularx}
 ~\newline
   \caption{Main sources and values of the relative systematic uncertainties of the \pt-differential yields for unidentified charged particles. The values are reported for low, intermediate and high \pt. The contributions that act differently in the various event classes are removed from the Total (quadratic sum of all contributions), defining the $N_{ch}$-independent ones, which are correlated across different multiplicity intervals.}
  \label{tab:sysunidentified}  
\end{table*}

\subsection{\label{subsecPiKP}Charged pions, kaons, and protons}

For the measurement of charged pions, kaons, and protons, several sub-analyses are combined for the comprehensive exploitation of the available PID techniques in ALICE. The spectra cover a range from 0.1/0.2/0.3 GeV/$c$ to 20 GeV/$c$ for \pipm/ \kapm/ \p($\rm\overline{p}$), respectively, with the exact ranges reported in Tab.~\ref{tab:OverviewIndividualAnalysesPiKP}. Similar approaches were followed in earlier analyses in \pp, \pPb\ and \PbPb\ collisions~\cite{Abelev:2013vea,Adam:2015qaa}.  An overview of the individual analyses is presented in Tab.~\ref{tab:OverviewIndividualAnalysesPiKP}. Here, we briefly review the most relevant aspects of previously employed techniques: ITS standalone and TPC-TOF. Additionally, we describe methods which are used for the first time for the measurement of \pt-spectra of charged kaons, pions, and protons: Bayesian PID, TPC-TOF fits, and TPC template fits.

In the ``ITS stand-alone'' technique, the average energy loss in the four outer ITS layers is calculated as a truncated mean. For each particle mass hypothesis, the distance between the measured and the expected value is calculated in multiples of the standard deviation $\sigma$ of the measured energy loss distribution and the particle mass hypothesis with the smallest value is assigned. In contrast to the analysis in the high track density environment of central heavy-ion collisions, the contribution of tracks with wrongly assigned signal clusters is negligible even in the highest pp event multiplicity class. 
In the intermediate \pt-range where a track-by-track identification is feasible, the ``TPC-TOF'' analysis identifies particles by requiring that the measured energy loss signals in the TPC and time-of-flight in the TOF are within 3$\sigma$ of the expected value assuming a specific mass hypothesis. This approach finds its natural limitation towards higher momenta, as the expected energy losses and flight times for different species are insufficiently different to allow for a clear separation. The \pt-ranges in which this procedure is applicable are given in Tab.~\ref{tab:OverviewIndividualAnalysesPiKP}. Two alternative methods, namely Bayesian PID and TPC-TOF fits, were employed in order to unfold the measured d$E$/d$x$ and TOF distributions. The Bayesian method of particle identification for the extraction of the minimum-bias spectra of pions, kaons, and protons is described in detail in~\cite{Adam:2016acv}. The {\it a priori} probabilities used in the Bayesian-approach analysis were extracted from the experimental data for the minimum bias event sample using an iterative procedure. The influence of different sets of {\it a priori} probabilities, determined for the lowest and highest event multiplicity bins, was evaluated and included in the systematic uncertainties. The actual identification of particles is based on the maximum probability method in which the most likely particle type is assigned to the track. 

\begin{table}[tbph!]
        \centering
        \begin{tabular}{cccccc}
\hline\hline
                Analysis & PID & \multicolumn{3}{c}{\pt\ range (\GeVc)} & (pseudo)rapidity\\
                \cline{3-5}
                         & technique                                                                         & \pipm            & \kapm         & p(\pbar)       & range\\ 
\hline
                \multirow{2}{*}{ITS stand-alone}  & n-$\sigma$ cuts  & \multirow{2}{*}{$0.1-0.6$}     & \multirow{2}{*}{$0.2-0.6$}   & \multirow{2}{*}{$0.3-0.6$}   & \multirow{2}{*}{$|y| < 0.5$}  \\ 
                                                  & on ITS & & & & \\
                                           &                                                        &                      &                   &                   &                    \\
                \multirow{2}{*}{Bayesian PID} & Bayesian & \multirow{2}{*}{$0.2-2.5$} & \multirow{2}{*}{$0.3-2.5$}   & \multirow{2}{*}{$0.5-2.5$}   & \multirow{2}{*}{$|y| < 0.5$}  \\
                                           & probability & & & & \\
                                           &                                                        &                      &                   &                   &                    \\
                \multirow{2}{*}{TPC-TOF}   & n-$\sigma$ cuts on & \multirow{2}{*}{$0.25-1.2$} & \multirow{2}{*}{$0.3-1.2$} & \multirow{2}{*}{$0.45-2.0$} & \multirow{2}{*}{$|y| < 0.5$}  \\
                                           & TPC and TOF & & & & \\
                                           &                                                        &                      &                   &                   &                    \\
                \multirow{2}{*}{TPC-TOF fits} & n-$\sigma$ fits    & \multirow{2}{*}{$0.25-2.5$}   & \multirow{2}{*}{$0.3-2.5$}   & \multirow{2}{*}{$0.45-2.7$} & $|y| < 0.5$ (TPC)\\
                                           &  to TPC and TOF     &  &  &  &$|\eta| < 0.2$ (TOF)\\
                                           &                                                        &                      &                   &                   &                    \\
                TPC template & TPC \dEdx        & \multicolumn{3}{c}{\multirow{2}{*}{$> 2.0$}}         & \multirow{2}{*}{$|\eta| < 0.8$}\\
                Fits   &   Template fits   & \multicolumn{3}{c}{} & \\
\hline\hline
        \end{tabular}
        ~\newline
                \caption{Overview of \pt\ ranges used for the combination of the various techniques used for identifying pions, kaons and protons. Since the true rapidity is not known at reconstruction level, fit based analyses (``TPC template fits" and ``TPC-TOF fits"), which determine the yield of pions, kaons and protons simultaneously, require an additional $\eta$-cut. }
        \label{tab:OverviewIndividualAnalysesPiKP}
\end{table}

In the ``TPC-TOF Fits'' method, the energy loss distribution in the TPC is simultaneously fitted by three gaussian distributions corresponding to charged pions, kaons and protons in each \pt\ and multiplicity bin. Similarly, the velocity distribution of the TOF is fitted for all three species simultaneously. In order to guarantee a sufficient separation of the particle species by minimizing the difference between total and transverse momentum \pt, the TOF fits were performed in a narrow $\eta$-window ($|\eta|<0.2$) and afterwards transformed to the common rapidity window of $|y|<0.5$ assuming a flat distribution in $y$.
Above a \pt\ of $\sim2$~GeV/$c$, particle identification can still be achieved statistically, rather than on a track-by-track basis, by fitting the specific energy loss in the relativistic rise region with a multi-component fit function, as done in the ``TPC Template fits'' approach. In this method, the measured d$E$/d$x$ distribution, in which the distributions of several particle species are overlapping, is fitted with a sum of templates (one for each particle type). The templates are extracted in a data-driven procedure from a pure sample of tagged particles of a given type. This pure sample is obtained from weak decay daughter tracks (\p\ and \pbar\ from $\lmb$ and $\almb$ as well as \pipm\ from \kzero) and tracks identified with the TOF (\pipm, \kapm, \p, and $\rm\overline{p}$). After a further strict selection of primary-particle-like topologies, the expected d$E$/d$x$ response is determined in fine bins of momentum and pseudorapidity. The template for each particle species in a given transverse momentum bin in the rapidity window $|y|<0.5$ is then obtained by sampling the measured momenta and pseudorapidity values of the tracks in this bin. The individual particle yields are the only free parameters in the fit of the templates to the measured d$E$/d$x$ distribution.

For all particle species and sub-analyses, contamination from secondary particles at low transverse momenta was subtracted in a data-driven approach on the basis of the measured distance of closest approach of the track to the primary vertex in the transverse plane (${\rm DCA_{xy}}$), as done in previous work~\cite{Adam:2015qaa}. The ${\rm DCA_{xy}}$ distribution of the selected tracks was fitted with three Monte Carlo templates corresponding to the expected shapes of primary particles, of secondaries from material (including electrons from photon conversions) and of secondaries from weak decays. The procedure was repeated for each \pt\ and event multiplicity bin and thus takes into account possible differences in the feed-down correction due to a change of the abundances and spectral shapes of weakly decaying strange particles. 

The efficiencies obtained for anti-protons and kaons have been additionally corrected based on a comparison of the absorption cross-section used in GEANT3 and the more realistic description of hadronic cross-sections in FLUKA, as in~\cite{Adam:2015qaa,Abbas:2013rua}. 

The determination of systematic uncertainties follows the procedures established in previous analyses~\cite{Abelev:2013vea,Adam:2015qaa}. All the considered contributions are summarized in Tab.~\ref{tab:syspikp}. Corrections for secondary particles lead to uncertainties of up to 4\% for protons and 1\% for pions while they are negligible for kaons. The uncertainty in the material budget is of 5\% at very low momenta and is related to the energy loss of the particles in the detector material. In addition, inelastic and elastic hadronic scattering processes inside the detector material are described by the transport codes only with limited precision and lead to uncertainties of up to 6\% for \pbar\ (for which the respective cross-section is largest) at low transverse momenta. The track quality selection criteria and the matching of TPC tracks with ITS hits give rise to a systematic uncertainty of the global tracking efficiency that amounts to 4\%, independent of \pt\ and particle species. 
The Lorentz force causes shifts of the cluster position in the ITS, pushing the charge in opposite directions depending on the polarity of the magnetic field of the experiment ($E \times B$ effect). In the ITS stand-alone analysis, the uncertainty related to this effect is estimated by analysing data samples collected with opposite magnetic field polarities, for which a difference of 3\% is observed.
For those sub-analyses (Bayesian PID, TPC-TOF, TPC-TOF fits) that require in addition that the track under study is matched to a hit in the TOF, an additional uncertainty of 3\%/6\%/4\% is taken into account for pions, kaons, and protons, respectively. Following the approach presented in~\cite{Abelev:2013vea}, this matching efficiency uncertainty was estimated by repeating the analysis separately for those regions in azimuth in which modules of the Transition Radiation Detector were already present in 2010 and for those in which they were not yet installed.

\begin{table*}[ph!]
  \begin{tabularx}{\textwidth}{p{5.3cm}*{3}{Y}*{3}{Y}*{3}{Y}}
\hline\hline
    &  \multicolumn{9}{c}{Uncertainty (\%)} \\
    {\bf Common source}  & \multicolumn{3}{c}{\pipm} & \multicolumn{3}{c}{\kapm} & \multicolumn{3}{c}{\p\ (\pbar)} \\ 
\hline
    \pt\ (\GeVc)   &   0.1 & 3.0 & 20.0   &   0.2 & 2.5 & 20.0 &   0.3 & 4.0 & 20.0   \\
    \cmidrule(l{5pt}){2-4} \cmidrule(l{5pt}){5-7} \cmidrule(l{5pt}){8-10} 
    Correction for secondaries & 1  & 1 & 1 & \multicolumn{3}{c}{negl.} & 4 & 1 & 1 \\
    Material budget & 5 & \multicolumn{2}{c}{negl.} & 2 & \multicolumn{2}{c}{negl.} & 4 & \multicolumn{2}{c}{negl.} \\
    Hadronic interactions & 2 & 1 & 1 & 3 & 1 & 1 & 4(6) & 1(1) & 1(1) \\
    Global tracking efficiency \newline (incl. track cut variation) & \multicolumn{3}{c}{\multirow{2}{*}{4}} & \multicolumn{3}{c}{\multirow{2}{*}{4}} & \multicolumn{3}{c}{\multirow{2}{*}{4}} \\
    TOF matching efficiency \newline (Bayes.,TPC-TOF,TPC-TOF fits) & \multicolumn{3}{c}{\multirow{2}{*}{3}} & \multicolumn{3}{c}{\multirow{2}{*}{6}} & \multicolumn{3}{c}{\multirow{2}{*}{4}} \\
\hline
    {\bf Specific source} & \multicolumn{3}{c}{\pipm} & \multicolumn{3}{c}{\kapm} & \multicolumn{3}{c}{\p\ (\pbar)} \\ 
\hline
    \pt\ (\GeVc)   &   0.1 & & 0.6   &   0.2 & & 0.6 &   0.3 & & 0.6   \\
    \cmidrule(l{5pt}){2-4} \cmidrule(l{5pt}){5-7} \cmidrule(l{5pt}){8-10} 
    ITSsa tracking efficiency & 3 &  & 3 & 3 &  & 3 & 3 &  & 3  \\ 
    $E \times B$ effect & \multicolumn{9}{c}{3} \\
    ITS PID & 5 & & 1 & 5 & & 9 & 8 & & 6 \\ 
\hline\hline

    \pt\ (\GeVc)   &   0.2 &  & 2.5   &   0.3 & & 2.5  &   0.5 & & 2.5    \\
    \cmidrule(l{5pt}){2-4} \cmidrule(l{5pt}){5-7} \cmidrule(l{5pt}){8-10} 
    Bayesian PID  & \multicolumn{3}{c}{1} & 1 & & 3 & 1 & & 2 \\   
\hline\hline
    
    \pt\ (\GeVc)   &   0.25 &  & 1.2   &   0.3 & & 1.2  &   0.45 & & 2.0    \\
    \cmidrule(l{5pt}){2-4} \cmidrule(l{5pt}){5-7} \cmidrule(l{5pt}){8-10} 
    TPC-TOF PID & \multicolumn{3}{c}{1} & 1 & & 5 & \multicolumn{3}{c}{1} \\
\hline\hline
    
    \pt\ (\GeVc)   &   0.25 &  & 2.5   &   0.3 & & 2.5  &   0.45 & & 2.7    \\
    \cmidrule(l{5pt}){2-4} \cmidrule(l{5pt}){5-7} \cmidrule(l{5pt}){8-10} 
    TPC-TOF fits PID  & 1 & & 5 & 1 & & 10 & 1 & & 8 \\
\hline\hline
    
    \pt\ (\GeVc)   &  2.0 &  & 20.0   &   2.0 & & 20.0  &   2.0 & & 20.0    \\
    \cmidrule(l{5pt}){2-4} \cmidrule(l{5pt}){5-7} \cmidrule(l{5pt}){8-10} 
    TPC template fits PID & 4 & & 6 & 10 & & 12  & 8 & & 13 \\
\hline\hline

   {\bf Total}  & \multicolumn{3}{c}{\pipm} & \multicolumn{3}{c}{\kapm} & \multicolumn{3}{c}{\p\ (\pbar)} \\ 
\hline
    \pt\ (\GeVc)   &   0.1 & 3.0 & 20.0   &   0.2 & 2.5 & 20.0 &   0.3 & 4.0 & 20.0   \\
    \cmidrule(l{5pt}){2-4} \cmidrule(l{5pt}){5-7} \cmidrule(l{5pt}){8-10} 
    Total & 8.4 & 5.0 & 7.2 & 7.5 & 6.6 & 12.6 & 12.3 & 15.1 & 13.3 \\
    Total ($N_{ch}$-independent) & 8.1 & 4.4 & 6.9 & 6.7 & 6.1 & 12.2 & 10.5 & 13.5 & 11.5\\
    \bottomrule
  \end{tabularx}
  ~\newline
    \caption{Main sources and values of the relative systematic uncertainties of the \pt-differential yields of \pipm, \kapm\ and \p (\pbar). The values are reported for low, intermediate and high \pt. The contributions that act differently in the various event classes are removed from the Total (quadratic sum of all contributions), defining the $N_{ch}$-independent ones, which are correlated across different multiplicity intervals. The contribution from the global tracking efficiency is common to all analyses except for the ITS stand-alone (ITSsa). 
}
  \label{tab:syspikp}  
\end{table*}

All sub-analyses were found to be in agreement in the overlapping \pt\ ranges within the uncorrelated part of their respective systematic uncertainties. The final combined spectrum for each particle species was then obtained by calculating the average over all sub-analyses using the uncorrelated part of their systematic errors as weights~\cite{Patrignani:2016xqp}. 
The uncertainties originating from common sources were then added in quadrature to each other and to the uncertainty attributed to the specific particle identification methods. The systematic uncertainty contribution that is uncorrelated across multiplicities was estimated to be $\sim$4-8\%, $\sim$6-12\% and $\sim$10-14\% for \pipm, \kapm and \p, respectively, for all V0M multiplicity bins.

\subsection{\label{subsecWeakDecay}Weakly-decaying Strange Hadrons}

The strange hadrons \kzero, \lmb, \almb, \X, \Ix, \Om\ and \Mo\ are reconstructed at mid-rapidity ($|y|<0.5$)
via their characteristic weak decay topology in the channels~\cite{Patrignani:2016xqp} 

\begin{center}
\begin{tabular*}{0.7\textwidth}{rll@{\extracolsep{2cm}}l}
  \kzero & $\to$ & \pPiplus + \pPiminus & {\footnotesize B.R. = (69.20 $\pm$ 0.05) \%} \\
  \lmb(\almb) & $\to$ & \pProton (\apProton) + \pPiminus (\pPiplus) & {\footnotesize B.R. = (63.9 $\pm$ 0.5) \%} \\
  \pXi (\apXi) & $\to$ & \pLambda (\apLambda) + \pPiminus (\pPiplus) & {\footnotesize B.R. = (99.887 $\pm$ 0.035) \%} \\ 
  \pOmega (\apOmega) & $\to$ & \pLambda (\apLambda) + \pKminus (\pKplus) & {\footnotesize B.R. = (67.8 $\pm$ 0.7) \%} \\
\end{tabular*}
\end{center}

Charged particle tracks are selected on the basis of compatibility of their energy loss in the TPC 
with the expected losses under the pion, kaon and proton mass hypotheses. They are then combined
 into weak decay candidates 
following the topology of a V-shaped decay for \kzero, \lmb\ and \almb\ (denoted `V0' decays) and a 
combination of a V0 decay and one additional charged track for \X, \Ix, \Om\ and \Mo
(denoted `cascade' decays). In addition to several geometrical criteria on the arrangement of
decay daughter tracks, \kzero, \lmb\ and \Om\ candidates are required to have a calculated 
mass that is incompatible with other species that decay in a similar topological arrangement, 
which are \lmb, \kzero\ and \X, respectively. This selection is commonly denoted `competing
decay rejection' and the exact numerical value depends on the invariant mass resolution 
for the competing particle species. Furthermore, candidates whose proper lifetimes
are unusually large for their expected species are also rejected to avoid 
combinatorial background from interactions with the detector material. 
The selection criteria used to define V0 and cascade decay candidates
are listed in Tab.~\ref{tab:V0Sels} and Tab.~\ref{tab:CascSels}, respectively. 

\begin{table}[h]
\centering
\begin{tabular}{ll}
\hline\hline
\noalign{\smallskip}
{\bf V0 selection criterion} & Value                   \\
\noalign{\smallskip}
\hline
\noalign{\smallskip}
DCA (h$^\pm$ to PV)                        & $>0.06$  cm                           \\ 
DCA (h$^-$ to h$^+$)                       & $<1.0$ standard deviations                  \\
Fiducial volume (R$_{2D}$)                 & $>$ 0.5 cm                    \\
V0 pointing angle                          & $\cos\theta_{V0}>$ 0.97 (0.995)                     \\
Proper Lifetime                         & $<$ 20 (30) cm/$c$                     \\
Competing V0 Rejection Window               &  $\pm$5 (10) \MeVmass                     \\
\noalign{\smallskip}
\hline\hline
\end{tabular}
~\newline
\caption{Selection criteria parameters utilized in the \kzero, \lmb\ and \almb\ analyses presented in
this work. If a criterion for \lmb\ and \kzero\ finding differs, the criterion for the \lmb\ hypothesis
is given in parentheses. The acronym 
DCA stands for ``distance of closest approach" and PV for ``primary event vertex". The pointing angle
$\theta$ is the angle between the momentum vector of the reconstructed V0, and the line
segment bound by the decay and primary vertices and R$_{2D}$ denotes the transverse distance from the 
detector center. 
\label{tab:V0Sels}
}
\end{table}

\begin{table}[h]
\centering
\begin{tabular}{ll}
\hline\hline
\noalign{\smallskip}
{\bf V0 from cascade selection criterion} & Value                      \\
\noalign{\smallskip}
\hline
\noalign{\smallskip}
DCA (baryon to PV)                         & $>0.03$ cm                           \\  
DCA (meson to PV)                          & $>0.04$ cm                           \\ 
DCA (h$^-$ to h$^+$)                       & $<1.5$ standard deviations                  \\
$\Lambda$ mass ($m_{\mathrm{V0}}$)         & $1.108<m_{\mathrm{V0}}<$ 1.124 \GeVmass     \\
Fiducial volume (R$_{2D}$)                 & $>$ 1.2 (1.1) cm                    \\
V0 pointing angle                          & $\cos(\theta_{\textrm{V0}})>$ 0.97                     \\

\noalign{\smallskip}
\hline\hline
\noalign{\smallskip}
{\bf Cascade finding criterion} & Value \\
\hline
\noalign{\smallskip} 
DCA (bachelor to PV)       & $>0.04$ cm\\
DCA (V0 to PV)                            & $>0.06$ cm\\
DCA ($\pi^{\pm}$ (K$^{\pm}$) to V0)       & $< 1.3$ cm\\    
Fiducial volume (R$_{2D}$)                & $>$ 0.6 (0.5) cm\\
Cascade pointing angle                    & $\cos(\theta_{\textrm{casc}})>$ 0.97\\
Proper Lifetime                  & $<$ 3$c\tau$ \\
Competing Cascade Rejection Window (\Oms\ only)                    & $\pm$8 \MeVmass\\
\noalign{\smallskip}                     
\hline\hline
\end{tabular}
~\newline
\caption{Selection criteria for V0 ($\Lambda$) from cascades, and cascades (\Xis\ and \Oms) presented in
this work. If a criterion for \Xis\ and \Oms\ finding differs, the criterion for \Oms\ hypothesis
is given in parentheses.
DCA stands for ``distance of closest approach" and PV for ``primary event vertex". The pointing angle
$\theta$ is the angle between the momentum vector of the reconstructed V0 or cascade and the line
segment bound by the decay and primary vertices  and R$_{2D}$ denotes the transverse distance from the 
detector center. 
The cascade track curvature is neglected, and $\tau$ refers to the average
lifetime for the two different cascade species. 
\label{tab:CascSels}
}
\end{table}

\begin{figure*}[h]
  \includegraphics[width=8cm,height=8cm]{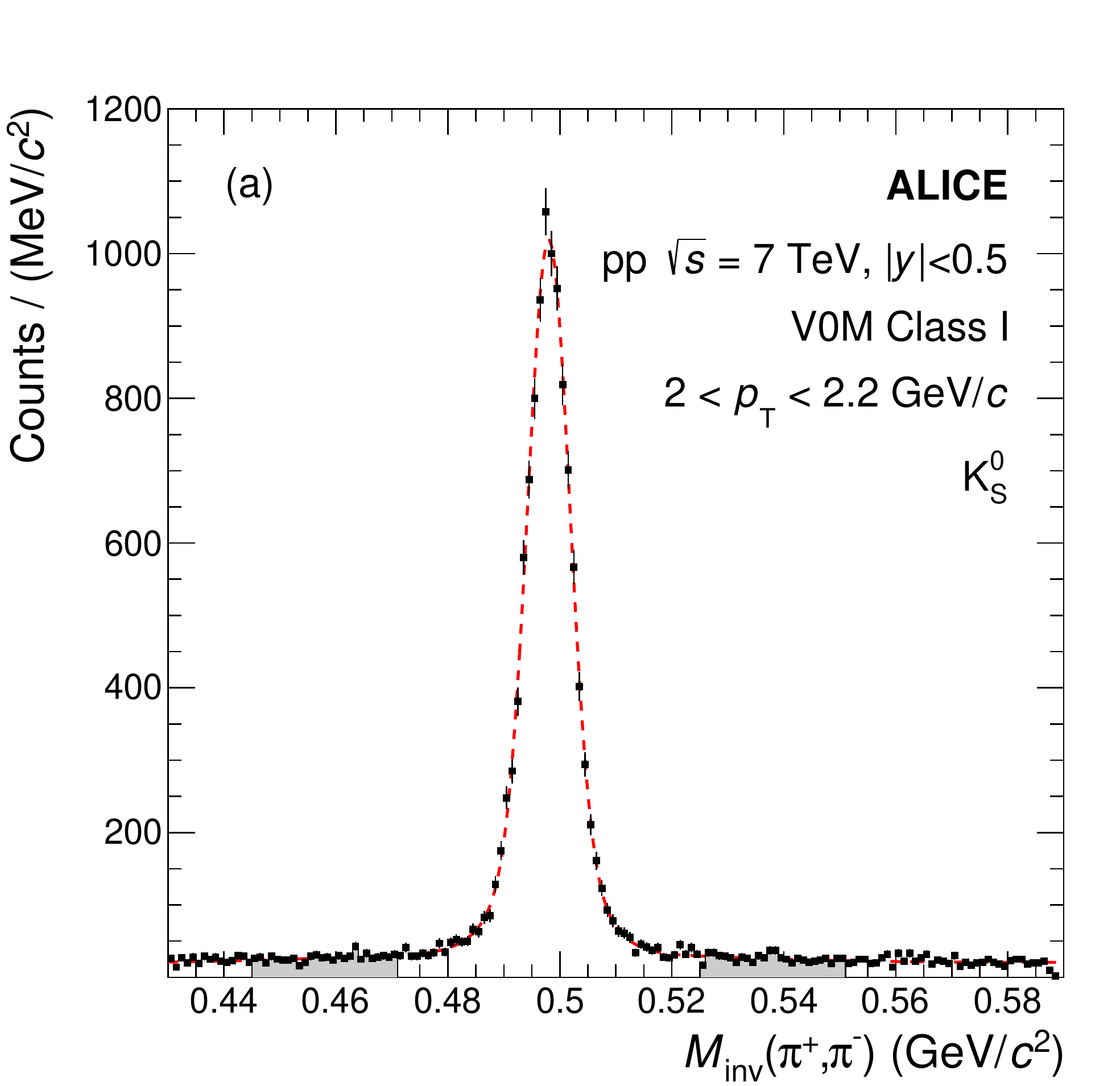}
  \includegraphics[width=8cm,height=8cm]{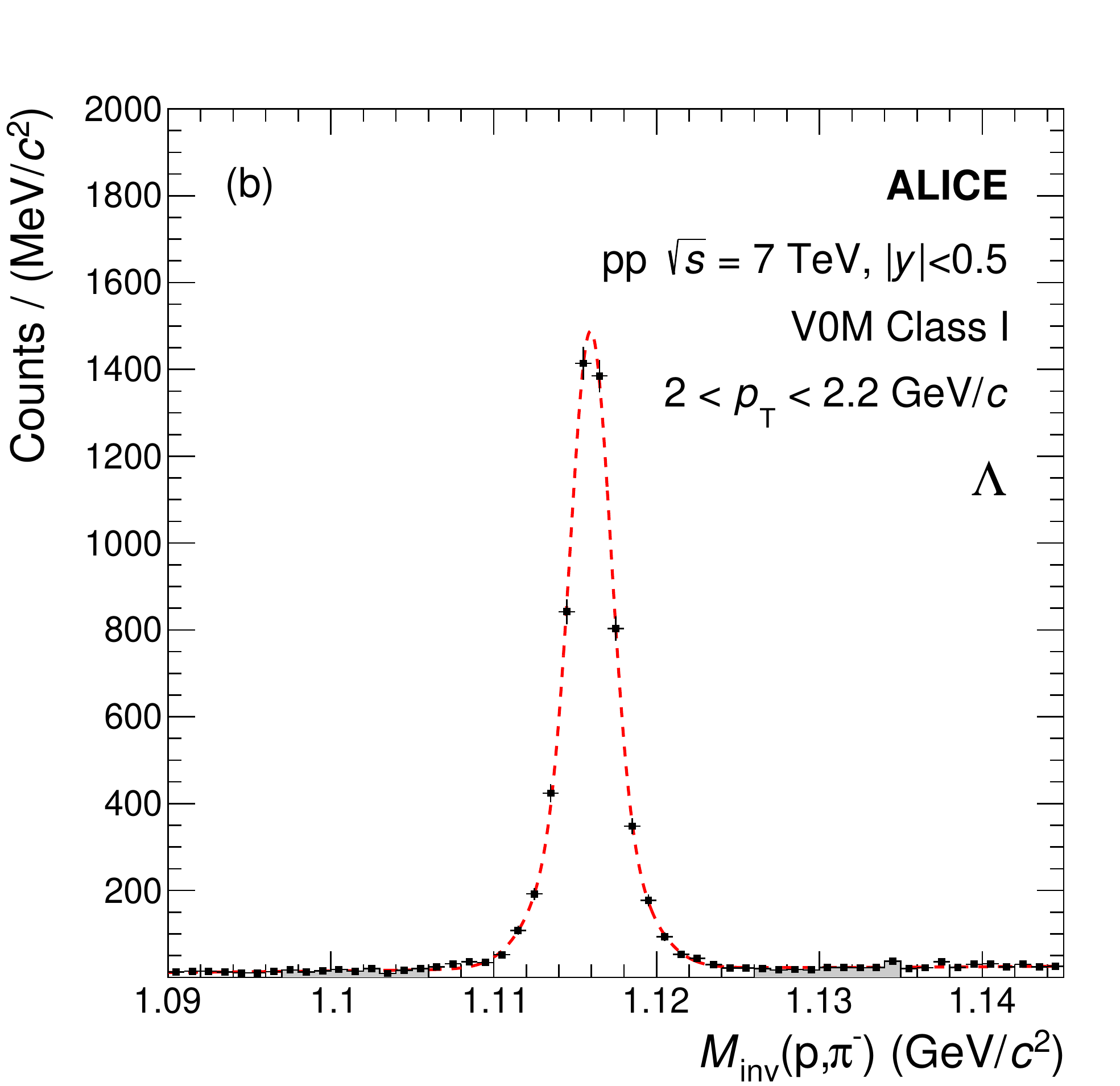}
  \\
  \includegraphics[width=8cm,height=8cm]{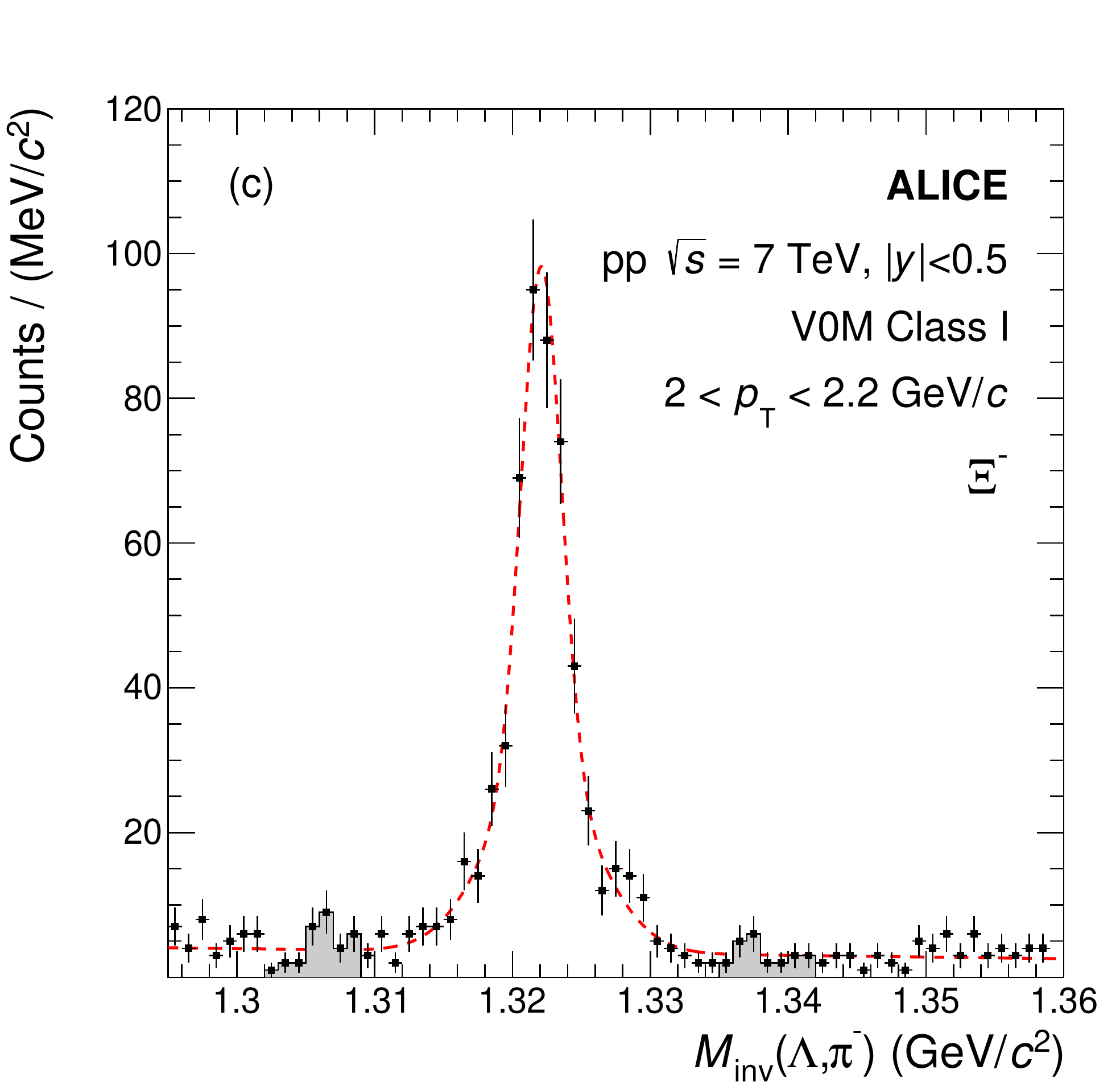}
  \includegraphics[width=8cm,height=8cm]{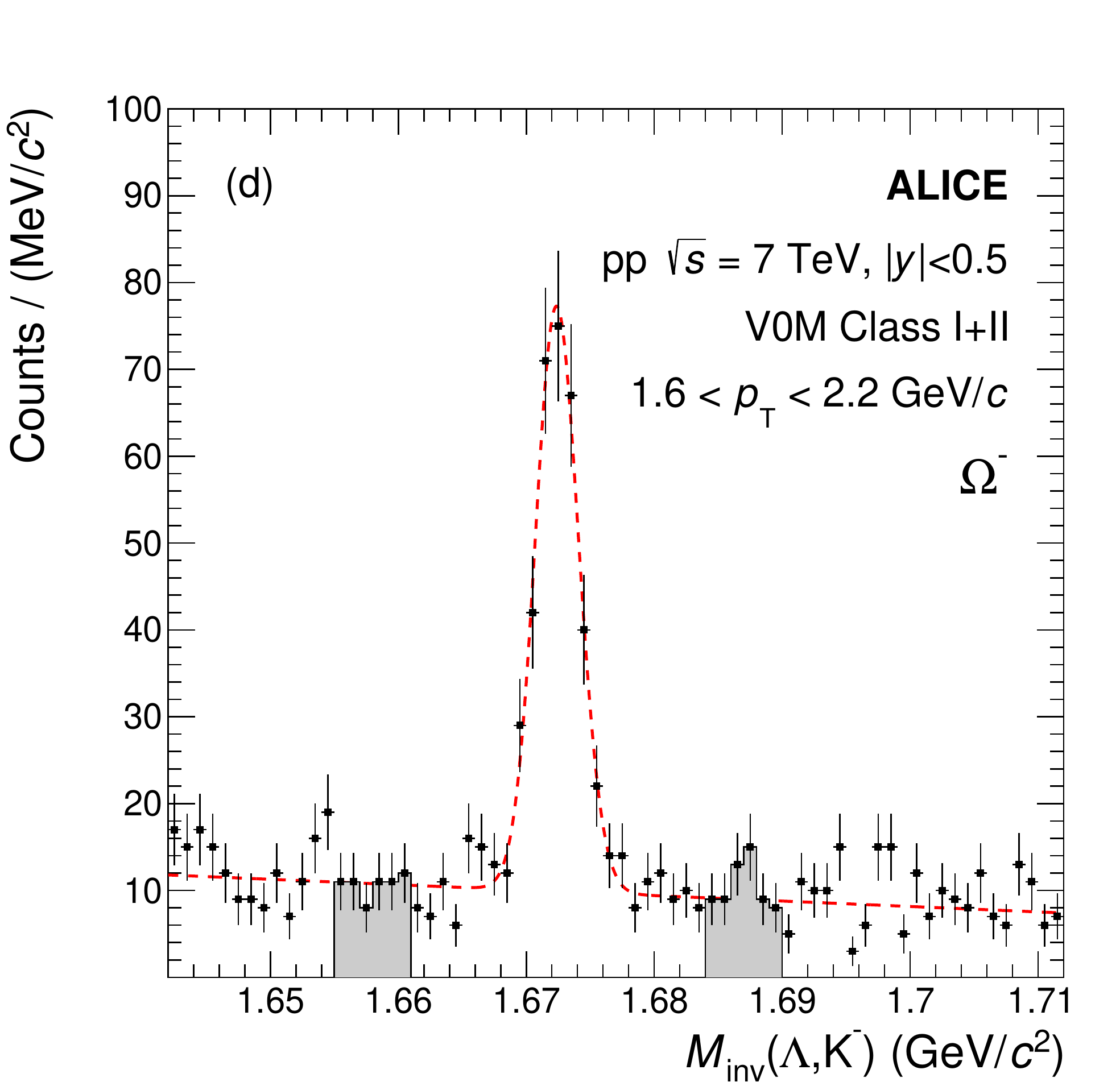}
  \caption{Invariant-mass distributions of (a) \kzero, (b) \lmb, (c) \X\ and (d) \Om\ (bottom right)  decay candidates in selected
    \pt\ ranges for the corresponding highest V0M event multiplicity classes in pp collisions at $\sqrt{s}$ = 7 TeV. The statistical uncertainties are shown by error bars
    and the shaded bands on the sides of the peak represent the regions used to estimate the background. The red dashed curves represent fits using a gaussian 
    peak and a linear background.}
  \label{fig:V0CascSig}
\end{figure*}

Particle yields are calculated in \pt\ and event multiplicity intervals by extracting the relevant signals
from invariant mass distributions as done in previous work~\cite{Abelev2012309, Abelev:2013haa, Adam:2015vsf}. 
Figure \ref{fig:V0CascSig} shows the invariant-mass distributions of \kzero\ (top left), \lmb\ (top right), \X\ (bottom left) 
and \Om\ (bottom right) in selected transverse momentum ranges for the corresponding highest V0M event multiplicity classes 
in pp collisions at $\sqrt{s}$ = 7 TeV.

\afterpage{\clearpage}

Approximately 20\% of the measured \lmb\ (\almb) signals are from \X\ (\Ix) and $\Xi^{0}$ ($\overline{\Xi}^{0}$)
decays. These feed-down contributions were subtracted using a data-driven approach in which the measured
\X\ (\Ix) spectra are used as input and a simulation is used to evaluate the fraction of reconstructed 
\lmb\ (\almb) coming from \X\ (\Ix) decays. Since production rates of $\Xi^{0}$ and $\overline{\Xi}^{0}$
 have not been measured, their contribution is estimated by assuming that they are as abundant as their
charged counterparts and that their momentum distributions are identical. 

Because in the specific case of the cascade analysis the measurement is performed in large momentum 
intervals because of the limited amount of data, efficiencies are reweighted in each \pt\ bin to take into 
account differences between generated and real data spectral shapes. 

Systematic uncertainties for \kzero, \lmb, \almb, \X, \Ix, \Om\ and \Mo\ are estimated following the 
procedure described in~\cite{Abelev:2013haa,  Adam:2015vsf}. The main sources of systematic 
uncertainty in these measurements are track selections (up to $\sim$6\%), knowledge of detector 
materials (4\%), feed-down from \X\ (\Ix) and $\Xi^{0}$ ($\overline{\Xi}^{0}$) for the \lmb\ (\almb) (up to $\sim$4\%) 
and topological selections, which contribute with a $\sim$1-8\% uncertainty. 
The contributions to 
systematic uncertainties are summarized in Tab.~\ref{tab:sysweakdecay}. As in previous work, 
the study of systematic uncertainties was repeated for all event classes to determine
differences in how each contribution affects results from each of these classes. 

\begin{table*}[ph!]
  \begin{tabularx}{\linewidth}{l*{3}{Y}*{3}{Y}}
\hline\hline
    Hadron species &  \multicolumn{3}{c}{\kzero} & \multicolumn{3}{c}{\lmb(\almb)} \\
    \pt\ range (\GeVc)                    &  0.05 & 6.2 & 11.0 &   0.5 & 3.7 & 7.2 \\
    \cmidrule(lr){2-4} \cmidrule(lr){5-7} 
    Material budget                 &   4.0 & 4.0 & 4.0  &   4.0 & 4.0 & 4.0 \\
    Transport code & \multicolumn{3}{c}{negligible}  &   1.0 & 1.0 & 1.0 \\
    Track selection                 &   1.0 & 5.0 & 0.8  &   0.2 & 5.9 & 4.3 \\ 
    Topological selection           &   2.6 & 1.1 & 2.3  &   0.8 & 0.6 & 3.2 \\ 
    Particle identification         &   0.1 & 0.1 & 0.1  &   0.2 & 0.2 & 3.0 \\ 
    Efficiency determination        &   2.0 & 2.0 & 2.0  &   2.0 & 2.0 & 2.0 \\
    Signal extraction               &   1.5 & 1.2 & 3.6  &   0.6 & 0.7 & 3.0 \\
    Proper lifetime                 &   1.3 & 0.1 & 0.2  &   0.3 & 2.3 & 0.1 \\
    Competing decay rejection       & negl. & 0.7 & 1.3  & negl. & 1.0 & 6.2 \\
    Feed-down correction & \multicolumn{3}{c}{not applicable}   &   3.3 & 2.1 & 4.3   \\
    \cmidrule(lr){2-4} \cmidrule(lr){5-7} 
    Total                           &   5.6 & 6.9 & 6.4  &   5.8 & 8.2 & 11.2 \\ 
    Common ($N_{ch}$-independent)   &   5.0 & 5.9 & 4.4  &   5.4 & 7.8 & 9.9  \\ 
\hline\hline
    Hadron species & \multicolumn{3}{c}{\X(\Ix)} & \multicolumn{3}{c}{\Om(\Mo)} \\
    \pt\ range (\GeVc)                    &   0.8 & 2.1 & 5.8   &   1.2 & 2.8 & 4.7 \\
    \cmidrule(lr){2-4} \cmidrule(lr){5-7} 
    Material budget                 &   4.0 & 4.0 & 4.0  &   4.0 & 4.0 & 4.0 \\
    Transport code                  &   1.0 & 1.0 & 1.0  &   1.0 & 1.0 & 1.0 \\
    Track selection                 &   0.4 & 0.3 & 2.2  &   0.8 & 0.6 & 4.1 \\ 
    Topological selection           &   3.1 & 2.0 & 4.0  &   5.0 & 5.6 & 8.1 \\ 
    Particle identification         &   1.0 & 0.2 & 1.2  &   1.1 & 1.7 & 3.2 \\ 
    Efficiency determination        &   2.0 & 2.0 & 2.0  &   2.0 & 2.0 & 2.0 \\
    Signal extraction               &   1.5 & 0.2 & 1.0  &   3.2 & 2.5 & 2.3 \\
    Proper lifetime                 &   0.9 & 0.1 & 0.1  &   2.2 & 0.7 & 0.7 \\
    Competing decay rejection   &  \multicolumn{3}{c}{not applicable}  &  0.2 & 4.2 & 5.2 \\
    Feed-down correction & \multicolumn{3}{c}{negligible}  &  \multicolumn{3}{c}{negligible} \\
    \cmidrule(lr){2-4} \cmidrule(lr){5-7} 
    Total                           &   5.9 & 5.0 & 6.7  &   7.9 & 9.0 & 12.1 \\ 
    Total ($N_{ch}$-independent)   &   5.2 & 4.5 & 6.2  &   7.3 & 8.7 & 11.6 \\ 
\hline\hline
  \end{tabularx}
  ~\newline
    \caption{Main sources and values of the relative systematic uncertainties (expressed in \%) of the \pt-differential yields of \kzero, \lmb(\almb), \X(\Ix) and \Om(\Mo). The values are reported for low, intermediate and high \pt. The contributions that act differently in the various event classes are removed from the Total (quadratic sum of all contributions), defining the $N_{ch}$-independent ones, which are correlated across different multiplicity intervals.}
  \label{tab:sysweakdecay}  
\end{table*}

\subsection{\label{subsecResonances}Resonances}

The \kstar\ and $\phi$-mesons are reconstructed at mid 
rapidity $|y|$ $<$ 0.5 via their hadronic decay channels into charged particles, 

\begin{center}
\begin{tabular*}{0.7\textwidth}{rll@{\extracolsep{2cm}}l}
  \kstar & $\to$ & $\pi^{\pm}$ + $K^{\mp}$ & {\footnotesize B.R. = ($\sim$66.6) \%} \\
  $\phi$ & $\to$ & $K^{+}$ + $K^{-}$ & {\footnotesize B.R. = (48.9 $\pm$ 0.5) \%} \\
\end{tabular*}
\end{center}

Both the TPC and TOF information are used to identify charged particles as 
pions or kaons from \kstar\ decays whereas only TPC information is used to identify charged particles as 
kaons from decays of $\phi$-mesons, as in the latter case the combinatorial background is significantly 
smaller.

\begin{figure*}[h]
    \includegraphics[width=8cm,height=8cm]{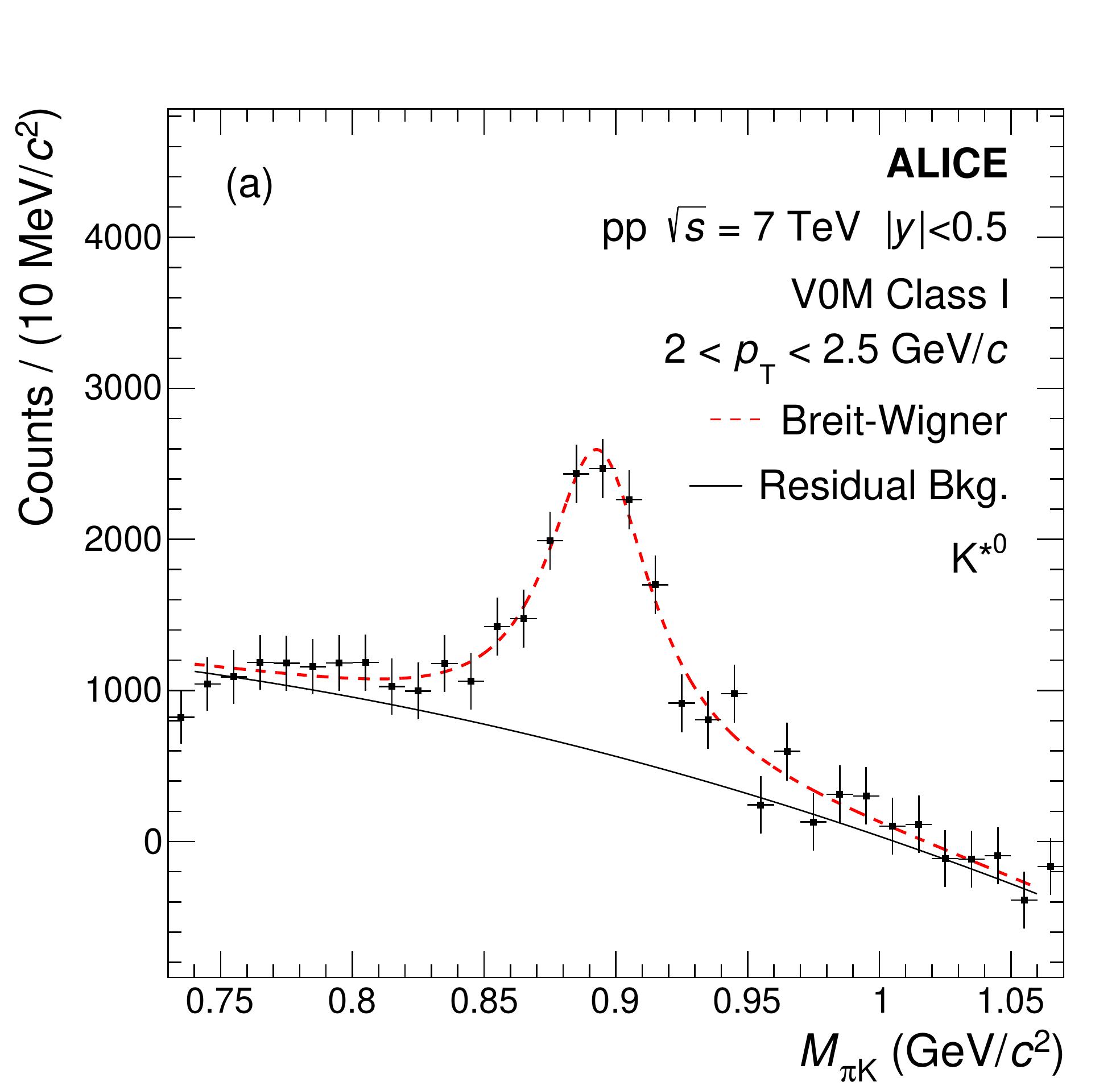}
    \includegraphics[width=8cm,height=8cm]{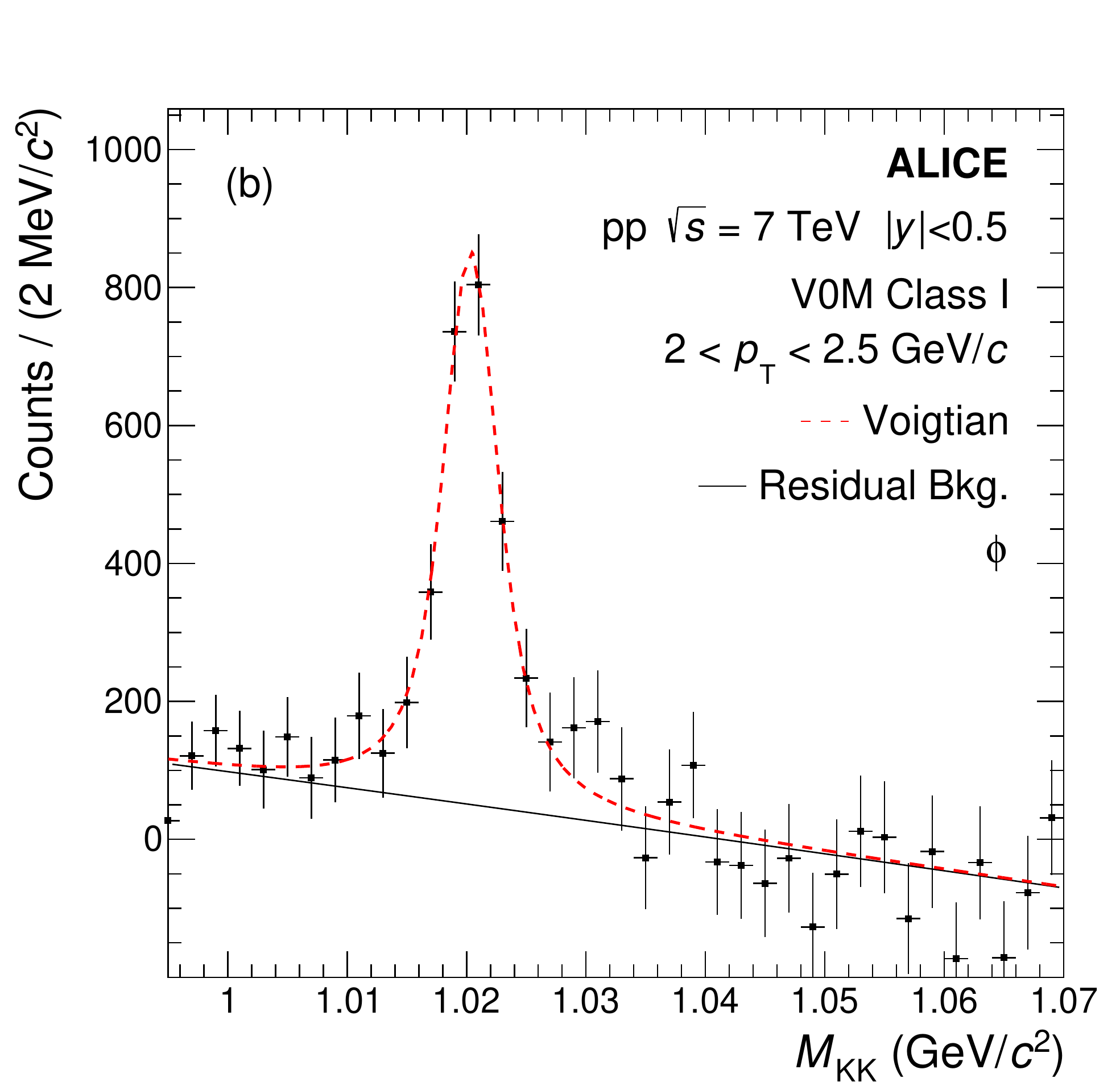}
  \caption{Invariant mass distributions of $\pi$K and KK in the momentum range of 2 $<$ \pt\ 
    $<$ 2.5 GeV/$c$ for V0M event multiplicity class I are 
    shown in panels (a) and (b), respectively. The statistical uncertainties are shown by vertical bars. The red dashed 
    curves represent fits to the distributions and the solid curves describe the residual
    background.}
  \label{fig:KstarPhiSig}
\end{figure*}

Pairs of pions and kaons (pairs of kaons) of opposite charge are considered to obtain the invariant mass distribution of \kstar\ ($\phi$) decay candidates. An event mixing 
technique is used to estimate the combinatorial background. The mixed-event distribution is 
normalized in the mass region outside of the mass peak, ${\it{i.e}}$ at 1.1 $<$ $M_{\pi\mathrm{K}}$ (GeV/$c^{2}$) 
$<$ 1.15 and 1.035 $<$ $M_{\mathrm{KK}}$ (GeV/$c^{2}$) $<$ 1.045  for \kstar\ and $\phi$-mesons, 
respectively. The normalized mixed-event distribution is subtracted from the same event unlike-sign 
distribution to isolate the relevant signals. 
After mixed-event background subtraction, each invariant mass distribution is fitted with a Breit-Wigner function (Voigtian function) for the signal and a 2nd-order polynomial for any residual background. The parameterizations for the signal are given in Eq.~\ref{kstarfitfun} for the \kstar\ and Eq.~\ref{phifitfun} for the $\phi$-meson:

\begin{equation}
      \frac{\mathrm{d}N}{\mathrm{d}M_{\pi \mathrm{K}}} = \frac{Y}{2\pi}
      \times \frac{\Gamma} {(M_{\pi \mathrm{K}} - M_0)^{2} + \frac{\Gamma^{2}}{4}}      
      \label{kstarfitfun}
\end{equation}

\begin{equation}
\frac{\mathrm{d}N}{\mathrm{d}M_\mathrm{KK}} = \frac{Y}{2\pi}\int \frac{\Gamma}
     {(M_ \mathrm{KK}-m^{'})^2+\Gamma^2/4}\times\frac{e^{{-(m^{'} -M_{0})}^{2}/2\sigma^2}}{\sqrt{2\pi}\sigma} dm^{'}
     \label{phifitfun}
\end{equation}

where $M_{\pi\mathrm{K}}$ and $M_{\mathrm{KK}}$ are the reconstructed invariant masses of \kstar\
and $\phi$-meson candidates, $M_{0}$, $\Gamma$ and $Y$ are the mass, width and raw yield of the resonances, respectively. The parameter $\sigma$ represents the mass resolution. Figure \ref{fig:KstarPhiSig} shows the invariant mass of $\pi$K (KK) in the left (right) panel 
for 2 $<$ \pt\ $<$ 2.5 GeV/$c$ in the V0M event multiplicity class I.

The raw yields are extracted in each \pt\ bin and event multiplicity interval as done in previous 
works~\cite{Abelev:2012hy,Adam:2016bpr,Abelev:2014uua}. In this analysis, detector 
acceptance and reconstruction efficiency are re-weighted in each 
\pt\ bin to take into account the differences between generated and real data spectral shapes.
  
The sources of systematic uncertainties for \kstar\ and $\phi$-meson production in pp collisions 
are the TPC-ITS matching efficiency, track selection criteria, PID, yield extraction method, hadronic 
interaction and material budget, and were evaluated following the same prescription used in previous 
works~\cite{Adam:2016bpr, Abelev:2012hy}. 
The main source of uncertainty for \kstar\ and $\phi$ comes from 
the determination of the TPC-ITS track matching efficiency. This contribution has been estimated 
to be a \pt-independent effect of 3$\%$ for charged particles~\cite{id2208submitted}, which 
results in a 6\% effect when 
any two primary tracks are combined in the invariant mass analysis of \kstar\ and $\phi$. For both \kstar\ 
and $\phi$, the uncertainties due to various track selection cuts from low to high-\pt\ are found 
to be 0.9-3.0\% and 1.6-2.4\%, respectively. The systematic uncertainty due to the signal extraction 
 includes variations in the fit range, fit function and normalization range and is of $\sim$5-10\% 
($\sim$3-9\%) from low to high \pt\ for \kstar\ ($\phi$). The uncertainty due to different PID selection methods
is estimated to be $\sim$2-4\% ($\sim$1-2$\%$) for  \kstar\ ($\phi$). The knowledge of the material budget 
for both \kstar\ and $\phi$ contributes to $\sim$4\% and $\sim$6\% at low \pt\ and is negligible at high \pt. 
The contribution from the estimate of the hadronic interaction cross section in the 
detector material at low-\pt\ is $\sim$4\% ($\sim$6\%) for \kstar\ ($\phi$) and negligible at high-\pt. The 
total systematic uncertainty for \kstar and $\phi$ is estimated to be about 12\% and 10\%, 
respectively. The maximum value of the multiplicity-independent systematic uncertainty is found 
to be $\sim$8\% ($\sim$5\%) for \kstar\ ($\phi$). The main contributions to the systematic 
uncertainties are summarized in Tab. \ref{tab:sysresonances}.
    
The systematic uncertainties were studied independently for all event classes, in order to separate the sources 
that are multiplicity-dependent and uncorrelated across multiplicity bins. In particular, signal 
extraction and PID are fully uncorrelated sources, whereas global tracking, track cuts, material 
budget and hadronic cross sections are correlated among different event multiplicity classes.

\begin{table*}[h]
  \begin{tabularx}{\linewidth}{l*{3}{Y}*{3}{Y}}
\hline\hline
\vspace{0.1cm}
    Hadron species &  \multicolumn{3}{c}{\kstar} & \multicolumn{3}{c}{$\phi$} \\
    \pt (\GeVc)       &   0.4 & 3.0 & 10.0 &   0.6 & 3.0 & 10.0  \\
    \cmidrule(lr){2-4} \cmidrule(lr){5-7} 
    Global tracking efficiency & \multicolumn{3}{c}{6} & \multicolumn{3}{c}{6} \\
    Signal extraction &5.1&   4.6&   9.7&   3.1&   3.2&   8.5 \\ 
    Track selection cuts &    3.0 &  2.1 & 0.9 &  1.6 &   1.6 &  2.4  \\
    Particle identification & 1.8 & 2.5 &  4.0 &   1.1  & 1.9  &  2.1 \\ 
    Material budget     & 4.3 &  0.8 &  0.1 &  6.2 &   0.4 &    - \\ 
    Hadronic interactions & 1.9 &0.9 & 0.1 & 1.4 & 0.7 & - \\
    \cmidrule(lr){2-4} \cmidrule(lr){5-7} 
    Total & 9.8&  8.3&  12.1&  9.5& 7.3&    9.0 \\ 
    Total ($N_{ch}$-independent) & 7.7&6.6&8.1 & 3& 5&5 \\ 
\hline\hline
  \end{tabularx}
  \newline
    \caption{Main sources and values of the relative systematic uncertainties (expressed in \%) of the \pt-differential yields of $\phi$ and \kstar\ resonances. The values are reported for low, intermediate and high \pt. The The contributions that act differently in the various event classes are removed from the total (quadratic sum of all contributions), defining the $N_{ch}$-independent ones, which are correlated across different multiplicity intervals. }
  \label{tab:sysresonances}  
\end{table*}

\begin{table*}[htb]
  \begin{tabularx}{\linewidth}{ X*{5}{c}} 
  \hline  \hline
  \multirow{ 2}{*}{Hadron species} & Lowest & Fraction of extrapolated & \multicolumn{2}{c}{Extrapolation uncertainty (\%)}  \\
\cmidrule(lr){4-5} 
& \pt\ (\GeVc) & {\dNdy\ (\%)} & {\dNdy} & {\meanpt\ (\GeVc)} \\
  \hline
  {\pipm} & 0.1 & {9$-$10} & {1.5} & {1.5} \\
  {\kapm} & 0.2 & {6$-$14} & {0.5$-$2} & {0.5$-$2} \\
  {\p\ (\pbar)} & 0.3 & {8$-$20} & {0.7$-$3} & {0.6$-$2.5} \\
  {\kzero} & 0.0 & {negl.} & {negl.} & {negl.} \\
  {\lmb(\almb)} & 0.4 & {10$-$25} & {2$-$6} & {2$-$4} \\
  {\X(\Ix)} & 0.6 &{16$-$36} &{3$-$10} &{2$-$7} \\
  {\Om(\Mo)} & 0.9 &{27$-$47} &{4$-$13} &{2$-$8} \\
 {$\phi$} & 0.4 &{10$-$24} & {2$-$5} &{2$-$4} \\
  {\kstar} & 0.0 &{negl.} & {negl.} &{negl.} \\
  \hline  \hline
  \end{tabularx}
    \newline
     \caption{Overview of the systematic uncertainties associated to the low-\pt\ extrapolation used to calculate \dNdy\ and \meanpt\ for the various particle species. Values for the highest and the lowest multiplicity classes are given, with the smallest and highest uncertainties being associated to higher and lower multiplicity classes, respectively.
If the dependence with multiplicity is negligible, only a single value is given.}
  \label{tab:extuncert} 
\end{table*}

\section{\label{secResults}Results}

\subsection{\label{ptspectra}Transverse momentum distributions} 

\begin{figure*}[htb!]
  \begin{flushleft}
    \center{\includegraphics[width=0.75\textwidth]{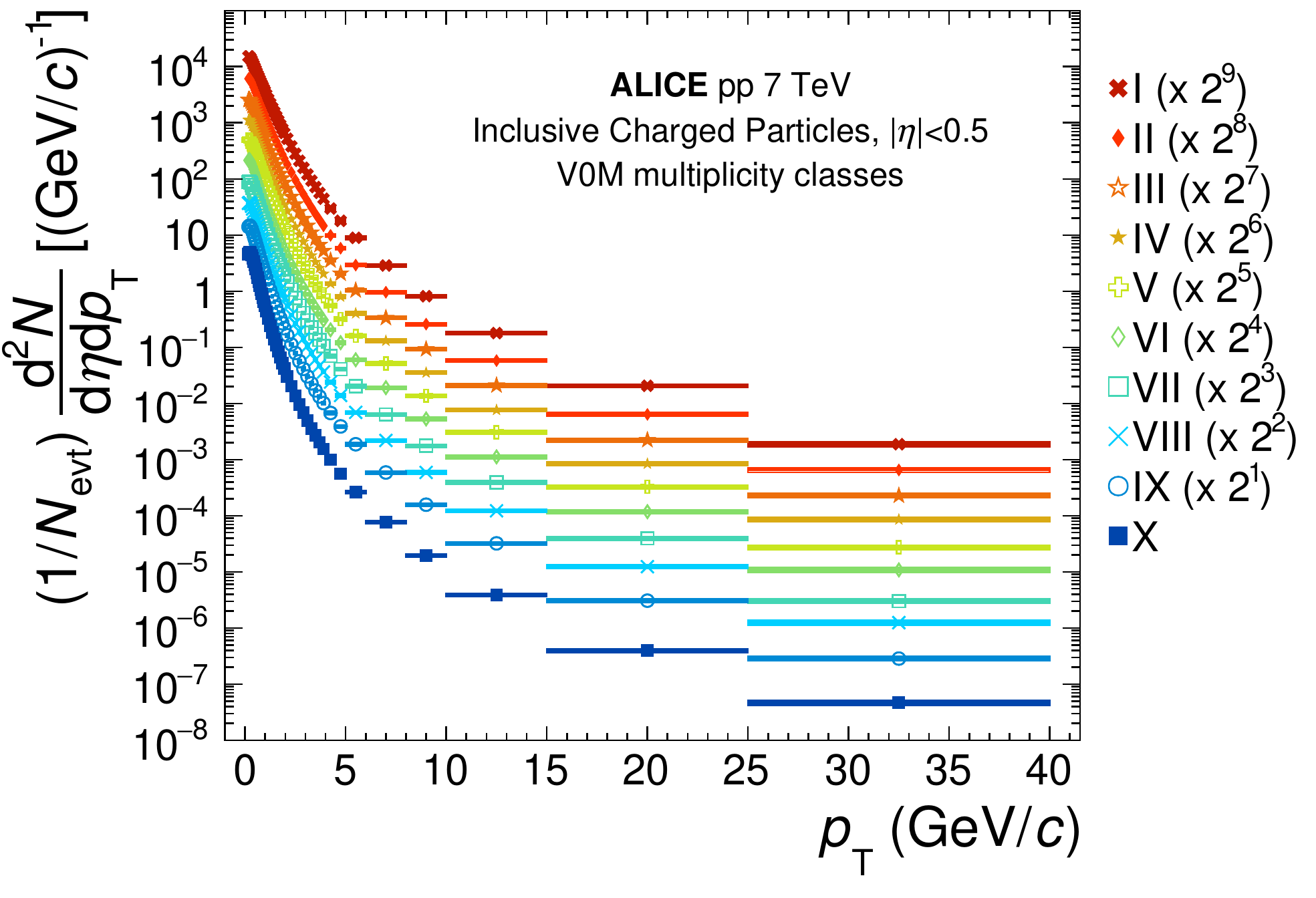}
    }
    \end{flushleft}
    \caption{Transverse momentum spectra of the sum of positively and negatively charged particles in different V0M event multiplicity classes.
} 
    \label{fig:UIDspectra}
\end{figure*}

\begin{figure*}[htb!]
    \center{\includegraphics[width=\textwidth]{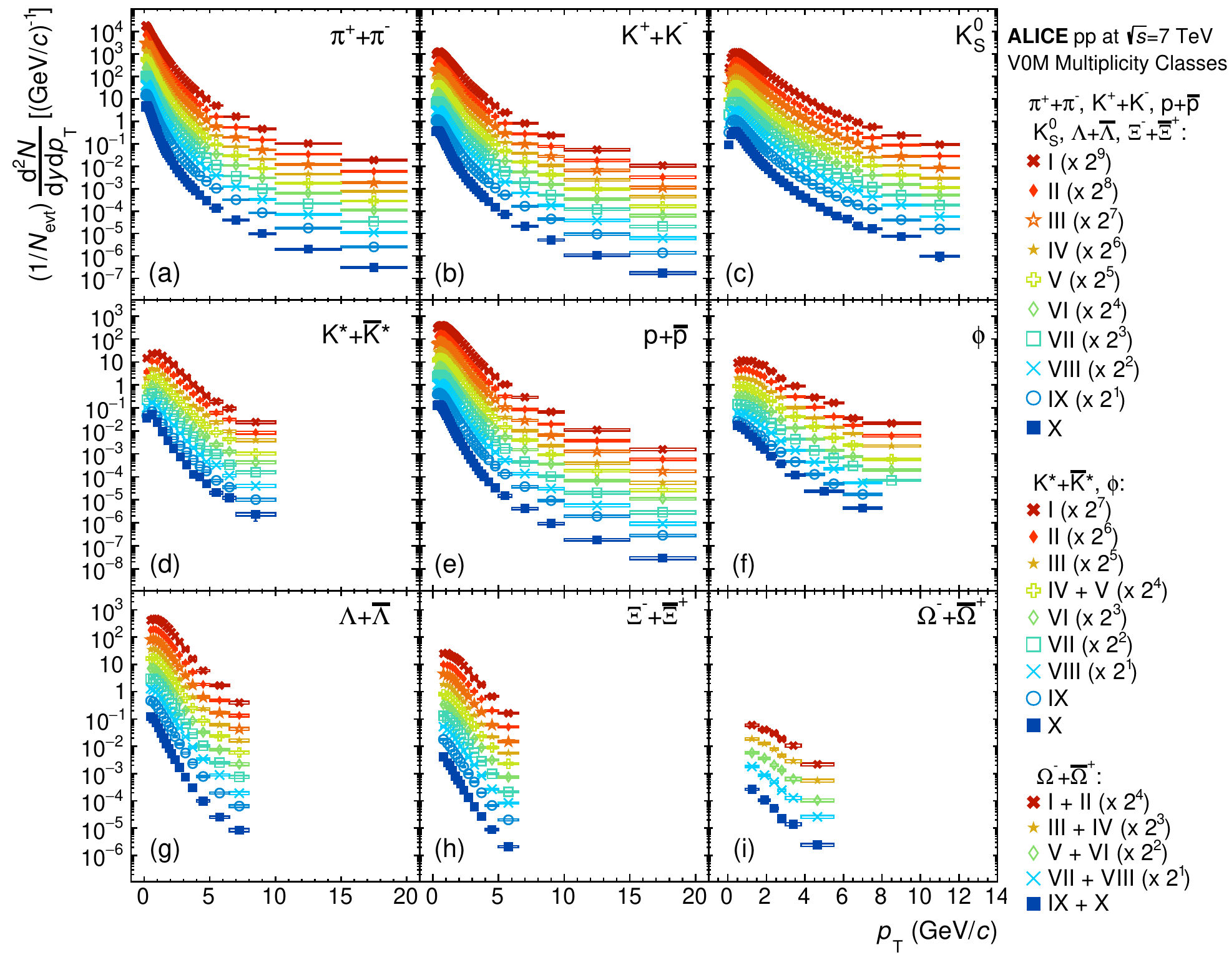}}
    \caption{Transverse momentum spectra of (a) \pipm, (b) \kapm, (c) \kzero, (d) \kstar, (e) \p+\pbar, (f) $\phi$, 
(g) $\lmb$+$\almb$, (h) \X+\Ix and (i) \Om+\Mo. Top to bottom: high to low multiplicity; data are scaled by $2^{n}$ factors for better visibility. }
    \label{fig:spectra}
\end{figure*}

The transverse momentum distributions measured at midrapidity for the 
event classes defined in Tab.~\ref{tab:OverviewEventClasses} are shown in Fig.~\ref{fig:UIDspectra}
for unidentified charged particles ($|\eta| < 0.5$) and Fig.~\ref{fig:spectra} for  \pipm, \kapm, \kzero, \kstar, \p, \pbar, $\phi$, 
$\lmb$, $\almb$, \X, \Ix, \Om\ and \Mo ($|y| < 0.5$). In the particular case of the $\phi$, \kstar\ and \Om\ measurements, some event classes were merged to allow for sufficient statistics. Particle and antiparticle as well as charged and neutral kaon production rates are compatible within uncertainties.

\begin{figure*}[htb!]
\center{\includegraphics[width=0.8\textwidth]{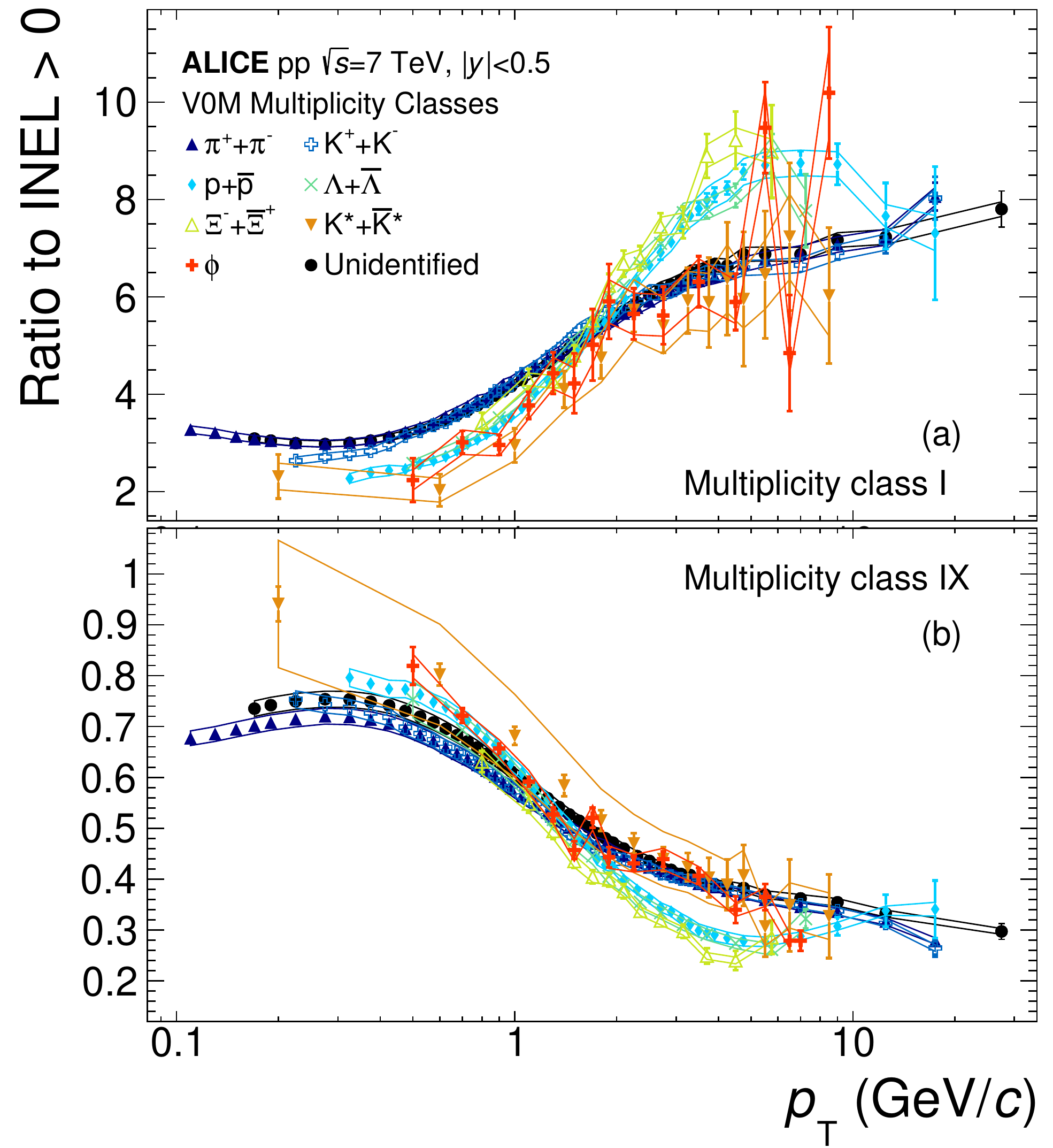}}
    \caption{Unidentified and identified particle spectra modification in (a) high-multiplicity and (b) low-multiplicity event classes. 
    Statistical uncertainties are shown as errors bars and systematic uncertainties that are uncorrelated across multiplicities are shown as boxes. Other 
    uncertainties are disregarded as they cancel in the ratio.}
    \label{fig:spectramodification}
\end{figure*}

Transverse momentum spectra are observed to become harder with increasing charged-particle multiplicity, with absolute changes in the spectrum shapes being more pronounced for particles with larger mass. The evolution of the \pt\ distributions with respect to the spectra in the INEL$>$0 event class for the various particle species is shown in Fig.~\ref{fig:spectramodification} and is observed to be identical for the two \pipm\ and \kapm\ mesons as well as for the \p, \lmb, \X\ baryons and their corresponding anti-particles. The spectra modification of $\phi$ and \kstar\ resonances follows the trend observed for baryons at \pt\ $<$ 2 \gevc\, while for larger momenta the modification is similar to the one observed for other mesons. Given that these mesonic resonances have a significantly higher mass than that of \pipm\ and \kapm, this suggests that the spectra evolution with multiplicity is driven by the hadronic mass at low \pt\ and by the number of constituent quarks at higher \pt. It is also interesting to note that such behavior 
is not unique to high multiplicity, but is present even for the lowest multiplicity class, where mass-dependent 
mechanisms such as radial flow are not expected to play a significant role. 

\begin{figure*}[tp!]
  \begin{flushleft}
    \center{\includegraphics[width=0.99\textwidth]{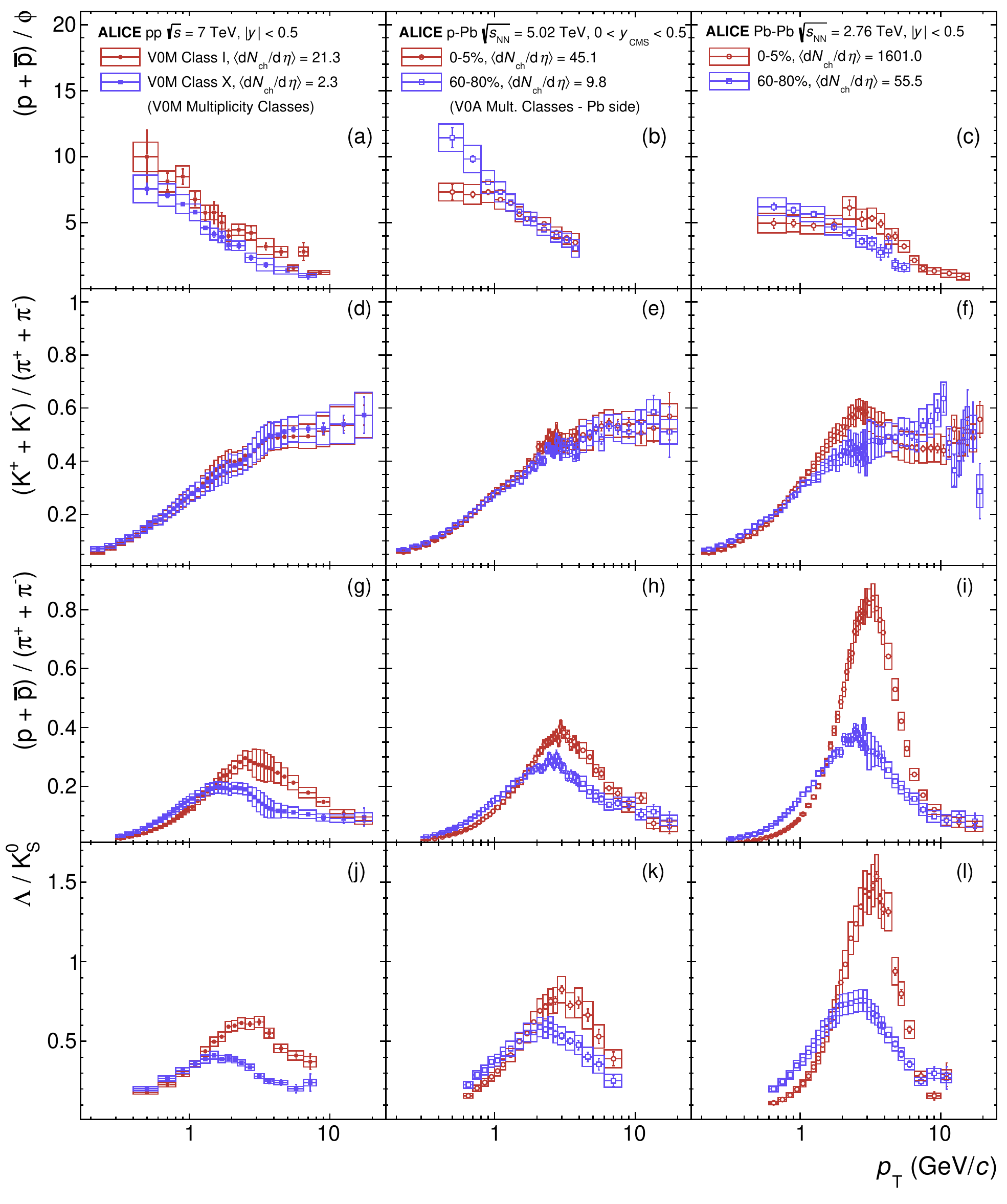}}

    \end{flushleft}
    \caption{Transverse momentum dependence of (a, b, c) \p/$\phi$~=~(p + \pbar)/$\phi$, (d, e, f) \kpi~=~(\kap + \kam)/(\pip + \pim),  (g, h, i) \ppi~=~(p + \pbar)/(\pip +
\pim) and (j, k, l) \lmb/\kzero\ (from top to bottom row) yield ratios in (a, d, g, j) pp, (b, e, h, k) p-Pb and (c, f, i, l) Pb-Pb collisions for high- and 
    low-multiplicity classes, respectively. Cancellation of common systematic uncertainties in the numerator and denominator was carried
    out only for the \lmb/\kzero, as in other cases the cancellation is non-trivial because of the use of various combined identification techniques or, in the case of resonances, of significantly different analysis strategy. Reference p-Pb and Pb-Pb data from \cite{Abelev:2013haa, Adam:2015vsf, 2016720, Abelev:2013xaa,2014196}.}
    \label{fig:RatiosVsPt}
\end{figure*}

A comparison of the multiplicity dependence of the transverse momentum distributions can be performed by studying the ratios \p/$\phi$~=~(p + \pbar)/$\phi$, \kpi~=~(\kap + \kam)/(\pip + \pim),  \ppi~=~(p + \pbar)/(\pip +
\pim) and  \lmb/\kzero\ as a function of
\pt, as shown in Fig.~\ref{fig:RatiosVsPt} for the lowest and highest multiplicity classes in this work. 
Results are compared with measurements in \pPb\ collisions at \fivenn~\cite{Abelev:2013haa, Adam:2015vsf, 2016720} as well as in \PbPb\ at 
\twosevensixnn~\cite{Abelev:2013xaa,2014196}. 
The \p/$\phi$ ratio is observed to decrease with \pt\ in \pp\ and 
low-multiplicity \pPb\ collisions but is seen to become progressively flatter for high-multiplicity \pPb\ and \PbPb\
collisions. Given the similar mass of the particles involved in this ratio, it is possible that this flattening is  
a signature of significant radial flow in the larger systems. 
Furthermore, the baryon-to-meson ratios  \ppi\ and \lmb/\kzero\ exhibit a characteristic depletion at \pt~$\sim0.7$~\gevc\ and an enhancement at intermediate \pt\ ($\sim3$~\gevc), which is qualitatively similar to that observed in \pPb\ and \PbPb\ collisions. Finally, the \kpi\ ratio is observed to be relatively
multiplicity-independent, except for central \PbPb\ collisions, where a weak depletion (enhancement) at
low (intermediate) \pt\ is visible. 

\begin{figure*}[tp!]
  \begin{flushleft}
    \center{\includegraphics[width=0.99\textwidth]{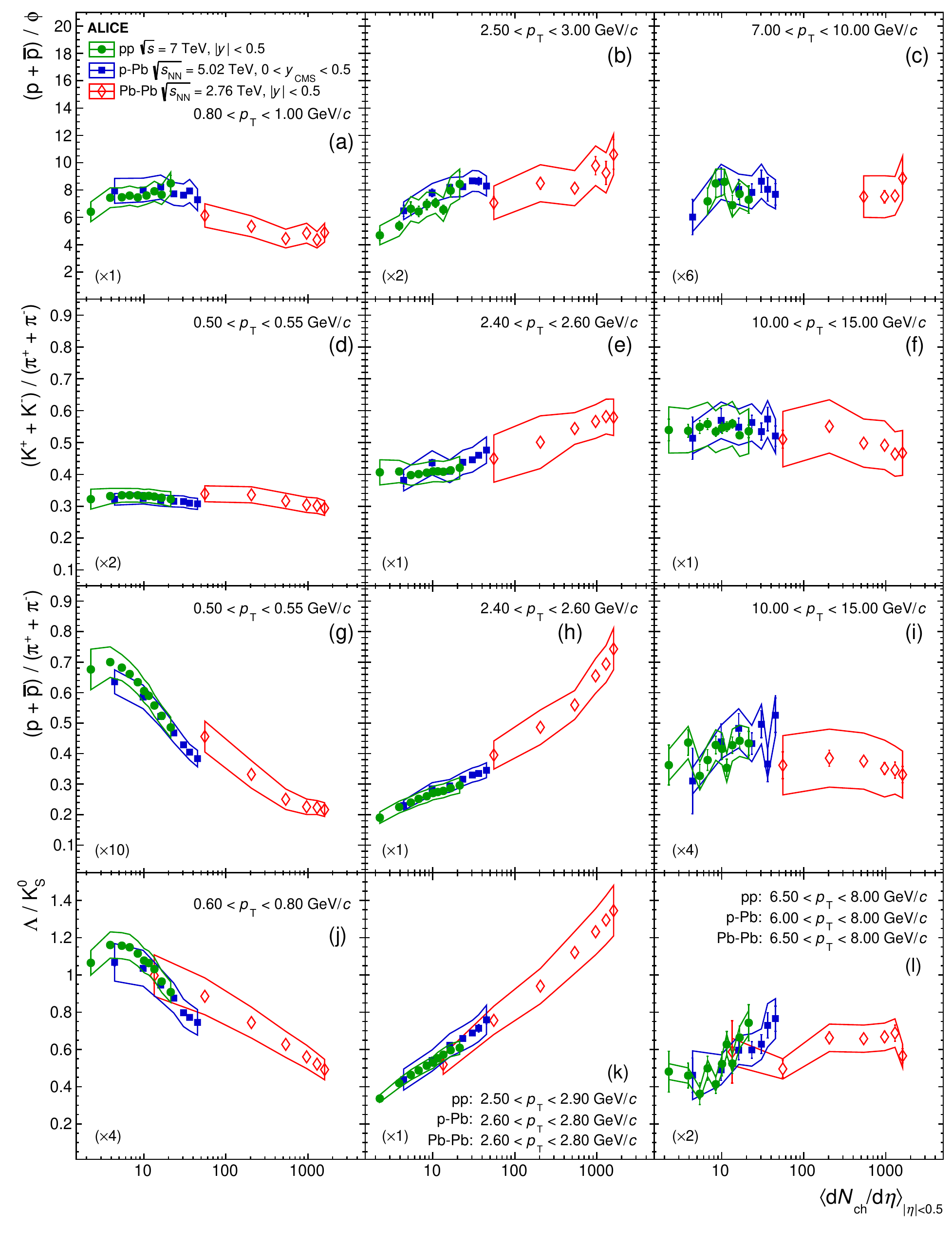}}
    \end{flushleft}
    \caption{Transverse momentum dependence of (a, b, c) \p/$\phi$~=~(p + \pbar)/$\phi$, (d, e, f) \kpi~=~(\kap + \kam)/(\pip + \pim),  (g, h, i) \ppi~=~(p + \pbar)/(\pip +
\pim) and (j, k, l) \lmb/\kzero\ yield ratios in (a, d, g, j) low-, (b, e, h, k) mid- and (c, f, i, l) high-\pt\ intervals 
in pp, p-Pb and Pb-Pb collisions as a function of \avdNdeta. Reference p-Pb and Pb-Pb data from \cite{Abelev:2013haa, Adam:2015vsf, 2016720, Abelev:2013xaa,2014196}.} 
    \label{fig:RatiosVsNch}
\end{figure*}

While the observed changes in these particle ratios are quantitatively different 
in the various collision systems, it is worth noting that the final state charged particle multiplicities also cover 
very distinct ranges. If considered as a function of \dNdeta, the ratios measured in specific low-, mid- and high-\pt\ intervals shown in Fig.~\ref{fig:RatiosVsNch} are seen to depend on multiplicity in a remarkably similar manner for all collision systems, despite differences in energy and collision geometry.

\clearpage

\subsection{\label{integrated}Integrated yields and average momenta} 

The \pt-integrated yields d$N$/d$y$ and mean transverse momenta \avpT\ have been calculated using data in the measured momentum range
 and a L\'{e}vy-Tsallis parametrization to extrapolate the spectra down to zero \pt, similarly to what 
was done in previous studies~\cite{Adam:2015qaa, Abelev2012309, Abelev:2012hy}. Several functional forms, such as
Boltzmann, \mt-exponential, \pt-exponential, Fermi-Dirac and Bose-Einstein functions, were used
to estimate the systematic uncertainties associated to this procedure. For those functions that are unable
to describe the \pt\ distributions in the full measured range, the fitting was restricted only to low-\pt. The uncertainty on \dNdy\ and \meanpt\ associated to this procedure is shown in Tab.~\ref{tab:extuncert}. 
The obtained \dNdy\ values for all multiplicity classes
under study are summarized in Tab.~\ref{tab:IntegratedYieldsTable} for identified particles, while the \avpT\ 
are reported in Tab.~\ref{tab:meanPtUnidentified} and \ref{tab:MeanPtTable} for unidentified charged and identified particles, 
respectively. 

\begin{table}[h]
\begin{center}
\noindent\makebox[\textwidth]{
\begin{tabular}{c|c}
\hline
V0M & \avpT\ (GeV/$c$) \\
\hline
I & 
(6.969 $\pm$ 0.001 $\pm$ 0.183)$\times10^{-1}$ \\
II & 
(6.672 $\pm$ 0.001 $\pm$ 0.176)$\times10^{-1}$ \\
III & 
(6.442 $\pm$ 0.001 $\pm$ 0.171)$\times10^{-1}$ \\
IV  & 
(6.275 $\pm$ 0.001 $\pm$ 0.167)$\times10^{-1}$ \\
V & 
(6.138 $\pm$ 0.001 $\pm$ 0.162)$\times10^{-1}$ \\
VI & 
(5.963 $\pm$ 0.001 $\pm$ 0.156)$\times10^{-1}$ \\
VII & 
(5.737 $\pm$ 0.001 $\pm$ 0.133)$\times10^{-1}$ \\
VIII & 
(5.527 $\pm$ 0.001 $\pm$ 0.106)$\times10^{-1}$ \\
IX & 
(5.214 $\pm$ 0.001 $\pm$ 0.103)$\times10^{-1}$ \\
X & 
(4.650 $\pm$ 0.001 $\pm$ 0.139)$\times10^{-1}$ \\
\hline
\end{tabular}
}
  ~\newline
\caption{\label{tab:meanPtUnidentified}Average transverse momenta ($|\eta| < 0.5$) for inclusive charged particles in different V0M event classes.}
\end{center}
\end{table}

\newcolumntype{M}[1]{>{\centering\arraybackslash}m{#1}}
\begin{sidewaystable}[p!] 
\caption{Integrated particle yields $\frac{\text{d}N}{\text{d}y}|_{|y|<0.5}$ for various V0M multiplicity classes. The first error represents the statistical uncertainty and the second one is the systematic and extrapolation errors added in quadrature. See text for details.}
\begin{center}
\begin{tabular}{ M{1cm} M{6cm} M{6cm} M{6cm} }
\hline\hline
\noalign{\smallskip
}V0M  &  $\frac{\pi^{+}+\pi^{-}}{2}$  &  $\frac{K^{+}+K^{-}}{2}$  &  $\frac{p+\overline{p}}{2}$ \\ 
\noalign{\smallskip}
\hline
I &10.035 $\pm$ 0.007 $\pm$ 0.519 & 1.374 $\pm$ 0.003 $\pm$ 0.093 & (5.488 $\pm$ 0.012 $\pm$ 0.393)$\times 10^{-1}$\\
II &7.878 $\pm$ 0.002 $\pm$ 0.377 & 1.062 $\pm$ 0.001 $\pm$ 0.067 & (4.369 $\pm$ 0.005 $\pm$ 0.301)$\times 10^{-1}$\\
III &6.459 $\pm$ 0.001 $\pm$ 0.306 & (8.590 $\pm$ 0.008 $\pm$ 0.539)$\times 10^{-1}$ & (3.599 $\pm$ 0.004 $\pm$ 0.246)$\times 10^{-1}$\\
IV &5.554 $\pm$ 0.001 $\pm$ 0.261 & (7.308 $\pm$ 0.006 $\pm$ 0.453)$\times 10^{-1}$ & (3.106 $\pm$ 0.003 $\pm$ 0.210)$\times 10^{-1}$\\
V &4.892 $\pm$ 0.001 $\pm$ 0.228 & (6.388 $\pm$ 0.006 $\pm$ 0.394)$\times 10^{-1}$ & (2.741 $\pm$ 0.003 $\pm$ 0.185)$\times 10^{-1}$\\
VI &4.138 $\pm$ 0.001 $\pm$ 0.192 & (5.337 $\pm$ 0.004 $\pm$ 0.327)$\times 10^{-1}$ & (2.316 $\pm$ 0.002 $\pm$ 0.156)$\times 10^{-1}$\\
VII &3.326 $\pm$ 0.001 $\pm$ 0.153 & (4.231 $\pm$ 0.003 $\pm$ 0.258)$\times 10^{-1}$ & (1.860 $\pm$ 0.002 $\pm$ 0.125)$\times 10^{-1}$\\
VIII &2.699 $\pm$ 0.001 $\pm$ 0.125 & (3.383 $\pm$ 0.003 $\pm$ 0.207)$\times 10^{-1}$ & (1.491 $\pm$ 0.002 $\pm$ 0.101)$\times 10^{-1}$\\
IX &(19.887 $\pm$ 0.003 $\pm$ 1.033)$\times 10^{-1}$ & (2.429 $\pm$ 0.002 $\pm$ 0.162)$\times 10^{-1}$ & (1.070 $\pm$ 0.001 $\pm$ 0.080)$\times 10^{-1}$\\
X &(12.100 $\pm$ 0.002 $\pm$ 0.857)$\times 10^{-1}$ & (1.402 $\pm$ 0.001 $\pm$ 0.115)$\times 10^{-1}$ & (5.856 $\pm$ 0.005 $\pm$ 0.532)$\times 10^{-2}$\\

\end{tabular}
\begin{tabular}{ M{1cm} M{6cm} M{6cm} M{6cm} }
\hline\hline
\noalign{\smallskip
}V0M  &  K$_{s}^{0}$  &  $\frac{\Lambda+\overline{\Lambda}}{2}$  &  K$^{*0}$+$\overline{\rm{K}}^{*0}$ \\ 
\noalign{\smallskip}
\hline
I &1.337 $\pm$ 0.004 $\pm$ 0.069 & (3.880 $\pm$ 0.014 $\pm$ 0.263)$\times 10^{-1}$ & (3.478 $\pm$ 0.185 $\pm$ 0.404)$\times 10^{-1}$\\
II &1.040 $\pm$ 0.002 $\pm$ 0.053 & (3.025 $\pm$ 0.007 $\pm$ 0.202)$\times 10^{-1}$ & (2.730 $\pm$ 0.078 $\pm$ 0.328)$\times 10^{-1}$\\
III &(8.415 $\pm$ 0.016 $\pm$ 0.430)$\times 10^{-1}$ & (2.445 $\pm$ 0.006 $\pm$ 0.165)$\times 10^{-1}$ & (2.348 $\pm$ 0.058 $\pm$ 0.293)$\times 10^{-1}$\\
IV &(7.195 $\pm$ 0.016 $\pm$ 0.366)$\times 10^{-1}$ & (2.076 $\pm$ 0.005 $\pm$ 0.140)$\times 10^{-1}$ & \multirow{2}{*}{(1.950 $\pm$ 0.045 $\pm$ 0.230)$\times 10^{-1}$}\\
V &(6.300 $\pm$ 0.015 $\pm$ 0.320)$\times 10^{-1}$ & (1.813 $\pm$ 0.005 $\pm$ 0.122)$\times 10^{-1}$ & {}\\
VI &(5.296 $\pm$ 0.010 $\pm$ 0.268)$\times 10^{-1}$ & (1.504 $\pm$ 0.004 $\pm$ 0.102)$\times 10^{-1}$ & (1.568 $\pm$ 0.027 $\pm$ 0.197)$\times 10^{-1}$\\
VII &(4.237 $\pm$ 0.009 $\pm$ 0.214)$\times 10^{-1}$ & (1.193 $\pm$ 0.003 $\pm$ 0.083)$\times 10^{-1}$ & (1.313 $\pm$ 0.033 $\pm$ 0.157)$\times 10^{-1}$\\
VIII &(3.393 $\pm$ 0.008 $\pm$ 0.171)$\times 10^{-1}$ & (9.361 $\pm$ 0.028 $\pm$ 0.616)$\times 10^{-2}$ & (1.081 $\pm$ 0.025 $\pm$ 0.129)$\times 10^{-1}$\\
IX &(2.462 $\pm$ 0.005 $\pm$ 0.124)$\times 10^{-1}$ & (6.450 $\pm$ 0.017 $\pm$ 0.461)$\times 10^{-2}$ & (8.088 $\pm$ 0.134 $\pm$ 1.070)$\times 10^{-2}$\\
X &(1.391 $\pm$ 0.003 $\pm$ 0.069)$\times 10^{-1}$ & (3.082 $\pm$ 0.010 $\pm$ 0.269)$\times 10^{-2}$ & (5.105 $\pm$ 0.073 $\pm$ 0.741)$\times 10^{-2}$\\

\end{tabular}
\begin{tabular}{ M{1cm} M{6cm} M{6cm} M{6cm} }
\hline\hline
\noalign{\smallskip
}V0M  &  $\phi$  &   $\frac{\Xi^{-}+\overline{\Xi}^{+}}{2}$  &  $\frac{\Omega^{-}+\overline{\Omega}^{+}}{2}$ 
\\ 
\noalign{\smallskip}
\hline
I &(1.697 $\pm$ 0.051 $\pm$ 0.172)$\times 10^{-1}$ & (4.808 $\pm$ 0.080 $\pm$ 0.304)$\times 10^{-2}$ & \multirow{2}{*}{(4.137 $\pm$ 0.132 $\pm$ 0.389)$\times 10^{-3}$}\\
II &(1.311 $\pm$ 0.021 $\pm$ 0.131)$\times 10^{-1}$ & (3.584 $\pm$ 0.037 $\pm$ 0.231)$\times 10^{-2}$ & {}\\
III &(1.098 $\pm$ 0.017 $\pm$ 0.114)$\times 10^{-1}$ & (2.966 $\pm$ 0.031 $\pm$ 0.200)$\times 10^{-2}$ & \multirow{2}{*}{(2.541 $\pm$ 0.098 $\pm$ 0.262)$\times 10^{-3}$}\\
IV &\multirow{2}{*}{(8.711 $\pm$ 0.104 $\pm$ 0.874)$\times 10^{-2}$} & (2.416 $\pm$ 0.027 $\pm$ 0.168)$\times 10^{-2}$ & {}\\
V &{} & (2.034 $\pm$ 0.025 $\pm$ 0.152)$\times 10^{-2}$ & \multirow{2}{*}{(1.488 $\pm$ 0.048 $\pm$ 0.149)$\times 10^{-3}$}\\
VI &(6.536 $\pm$ 0.088 $\pm$ 0.673)$\times 10^{-2}$ & (1.666 $\pm$ 0.018 $\pm$ 0.127)$\times 10^{-2}$ & {}\\
VII &(5.178 $\pm$ 0.076 $\pm$ 0.536)$\times 10^{-2}$ & (1.243 $\pm$ 0.015 $\pm$ 0.102)$\times 10^{-2}$ & \multirow{2}{*}{(9.313 $\pm$ 0.589 $\pm$ 1.186)$\times 10^{-4}$}\\
VIII &(3.988 $\pm$ 0.068 $\pm$ 0.412)$\times 10^{-2}$ & (9.443 $\pm$ 0.142 $\pm$ 0.849)$\times 10^{-3}$ & {}\\
IX &(2.905 $\pm$ 0.042 $\pm$ 0.308)$\times 10^{-2}$ & (6.023 $\pm$ 0.087 $\pm$ 0.638)$\times 10^{-3}$ & \multirow{2}{*}{(2.883 $\pm$ 0.301 $\pm$ 0.525)$\times 10^{-4}$}\\
X &(1.729 $\pm$ 0.029 $\pm$ 0.221)$\times 10^{-2}$ & (2.599 $\pm$ 0.053 $\pm$ 0.373)$\times 10^{-3}$ & {}\\

\hline\hline

\end{tabular}
\label{tab:IntegratedYieldsTable}
\end{center}
\end{sidewaystable}

\newcolumntype{M}[1]{>{\centering\arraybackslash}m{#1}}
\begin{sidewaystable}[p!]
\caption{Mean $p_{\text{T}}$ (GeV/$c$) for various V0M multiplicity classes. The first error represents the statistical uncertainty and the second one is the systematic. See text for details.}
\begin{center}
\begin{tabular}{ M{1cm} M{6cm} M{6cm} M{6cm} }
\hline\hline
\noalign{\smallskip
}V0M  &  $\frac{\pi^{+}+\pi^{-}}{2}$  &  $\frac{K^{+}+K^{-}}{2}$  &  $\frac{p+\overline{p}}{2}$ \\ 
\noalign{\smallskip}
\hline
I &(5.359 $\pm$ 0.003 $\pm$ 0.120)$\times 10^{-1}$ & (9.332 $\pm$ 0.009 $\pm$ 0.142)$\times 10^{-1}$ & 1.135 $\pm$ 0.001 $\pm$ 0.022\\
II &(5.158 $\pm$ 0.001 $\pm$ 0.108)$\times 10^{-1}$ & (8.892 $\pm$ 0.004 $\pm$ 0.127)$\times 10^{-1}$ & 1.071 $\pm$ 0.001 $\pm$ 0.021\\
III &(5.008 $\pm$ 0.001 $\pm$ 0.104)$\times 10^{-1}$ & (8.558 $\pm$ 0.004 $\pm$ 0.135)$\times 10^{-1}$ & 1.020 $\pm$ 0.001 $\pm$ 0.021\\
IV &(4.898 $\pm$ 0.001 $\pm$ 0.100)$\times 10^{-1}$ & (8.317 $\pm$ 0.004 $\pm$ 0.131)$\times 10^{-1}$ & (9.815 $\pm$ 0.007 $\pm$ 0.197)$\times 10^{-1}$\\
V &(4.806 $\pm$ 0.001 $\pm$ 0.095)$\times 10^{-1}$ & (8.113 $\pm$ 0.004 $\pm$ 0.133)$\times 10^{-1}$ & (9.539 $\pm$ 0.007 $\pm$ 0.200)$\times 10^{-1}$\\
VI &(4.693 $\pm$ 0.001 $\pm$ 0.091)$\times 10^{-1}$ & (7.867 $\pm$ 0.003 $\pm$ 0.130)$\times 10^{-1}$ & (9.170 $\pm$ 0.006 $\pm$ 0.195)$\times 10^{-1}$\\
VII &(4.557 $\pm$ 0.001 $\pm$ 0.086)$\times 10^{-1}$ & (7.538 $\pm$ 0.003 $\pm$ 0.133)$\times 10^{-1}$ & (8.725 $\pm$ 0.006 $\pm$ 0.196)$\times 10^{-1}$\\
VIII &(4.427 $\pm$ 0.001 $\pm$ 0.082)$\times 10^{-1}$ & (7.231 $\pm$ 0.004 $\pm$ 0.133)$\times 10^{-1}$ & (8.285 $\pm$ 0.006 $\pm$ 0.192)$\times 10^{-1}$\\
IX &(4.237 $\pm$ 0.001 $\pm$ 0.081)$\times 10^{-1}$ & (6.755 $\pm$ 0.003 $\pm$ 0.149)$\times 10^{-1}$ & (7.716 $\pm$ 0.005 $\pm$ 0.222)$\times 10^{-1}$\\
X &(3.904 $\pm$ 0.001 $\pm$ 0.090)$\times 10^{-1}$ & (5.885 $\pm$ 0.003 $\pm$ 0.159)$\times 10^{-1}$ & (6.647 $\pm$ 0.004 $\pm$ 0.220)$\times 10^{-1}$\\

\end{tabular}
\begin{tabular}{ M{1cm} M{6cm} M{6cm} M{6cm} }
\hline\hline
\noalign{\smallskip
}V0M  &  K$_{s}^{0}$  &  $\frac{\Lambda+\overline{\Lambda}}{2}$  &  K$^{*0}$+$\overline{\rm{K}}^{*0}$ \\ 
\noalign{\smallskip}
\hline
I &(9.367 $\pm$ 0.025 $\pm$ 0.119)$\times 10^{-1}$ & 1.277 $\pm$ 0.003 $\pm$ 0.028 & 1.307 $\pm$ 0.038 $\pm$ 0.043\\
II &(8.915 $\pm$ 0.015 $\pm$ 0.114)$\times 10^{-1}$ & 1.201 $\pm$ 0.002 $\pm$ 0.026 & 1.295 $\pm$ 0.022 $\pm$ 0.035\\
III &(8.612 $\pm$ 0.014 $\pm$ 0.108)$\times 10^{-1}$ & 1.144 $\pm$ 0.002 $\pm$ 0.026 & 1.211 $\pm$ 0.018 $\pm$ 0.037\\
IV &(8.350 $\pm$ 0.014 $\pm$ 0.104)$\times 10^{-1}$ & 1.103 $\pm$ 0.002 $\pm$ 0.025 & \multirow{2}{*}{1.150 $\pm$ 0.016 $\pm$ 0.035}\\
V &(8.155 $\pm$ 0.015 $\pm$ 0.101)$\times 10^{-1}$ & 1.069 $\pm$ 0.002 $\pm$ 0.025 & {}\\
VI &(7.900 $\pm$ 0.011 $\pm$ 0.098)$\times 10^{-1}$ & 1.027 $\pm$ 0.001 $\pm$ 0.025 & 1.064 $\pm$ 0.011 $\pm$ 0.033\\
VII &(7.563 $\pm$ 0.014 $\pm$ 0.094)$\times 10^{-1}$ & (9.759 $\pm$ 0.015 $\pm$ 0.242)$\times 10^{-1}$ & 1.014 $\pm$ 0.015 $\pm$ 0.031\\
VIII &(7.299 $\pm$ 0.014 $\pm$ 0.089)$\times 10^{-1}$ & (9.309 $\pm$ 0.016 $\pm$ 0.221)$\times 10^{-1}$ & (9.557 $\pm$ 0.135 $\pm$ 0.292)$\times 10^{-1}$\\
IX &(6.869 $\pm$ 0.012 $\pm$ 0.082)$\times 10^{-1}$ & (8.708 $\pm$ 0.013 $\pm$ 0.255)$\times 10^{-1}$ & (8.656 $\pm$ 0.088 $\pm$ 0.283)$\times 10^{-1}$\\
X &(6.113 $\pm$ 0.011 $\pm$ 0.068)$\times 10^{-1}$ & (7.669 $\pm$ 0.014 $\pm$ 0.319)$\times 10^{-1}$ & (7.185 $\pm$ 0.058 $\pm$ 0.245)$\times 10^{-1}$\\

\end{tabular}
\begin{tabular}{ M{1cm} M{6cm} M{6cm} M{6cm} }
\hline\hline
\noalign{\smallskip
}V0M  &  $\phi$  &   $\frac{\Xi^{-}+\overline{\Xi}^{+}}{2}$  &  $\frac{\Omega^{-}+\overline{\Omega}^{+}}{2}$ 
\\ 
\noalign{\smallskip}
\hline
I &1.440 $\pm$ 0.023 $\pm$ 0.037 & 1.463 $\pm$ 0.012 $\pm$ 0.036 & \multirow{2}{*}{1.645 $\pm$ 0.025 $\pm$ 0.056}\\
II &1.360 $\pm$ 0.012 $\pm$ 0.034 & 1.382 $\pm$ 0.007 $\pm$ 0.035 & {}\\
III &1.297 $\pm$ 0.011 $\pm$ 0.036 & 1.293 $\pm$ 0.007 $\pm$ 0.037 & \multirow{2}{*}{1.561 $\pm$ 0.033 $\pm$ 0.071}\\
IV &\multirow{2}{*}{1.248 $\pm$ 0.009 $\pm$ 0.030} & 1.268 $\pm$ 0.007 $\pm$ 0.037 & {}\\
V &{} & 1.232 $\pm$ 0.008 $\pm$ 0.042 & \multirow{2}{*}{1.465 $\pm$ 0.022 $\pm$ 0.065}\\
VI &1.201 $\pm$ 0.010 $\pm$ 0.034 & 1.188 $\pm$ 0.006 $\pm$ 0.044 & {}\\
VII &1.127 $\pm$ 0.010 $\pm$ 0.033 & 1.141 $\pm$ 0.007 $\pm$ 0.048 & \multirow{2}{*}{1.283 $\pm$ 0.047 $\pm$ 0.074}\\
VIII &1.081 $\pm$ 0.011 $\pm$ 0.031 & 1.092 $\pm$ 0.008 $\pm$ 0.048 & {}\\
IX &(9.910 $\pm$ 0.087 $\pm$ 0.268)$\times 10^{-1}$ & 1.012 $\pm$ 0.007 $\pm$ 0.054 & \multirow{2}{*}{1.125 $\pm$ 0.061 $\pm$ 0.101}\\
X &(8.225 $\pm$ 0.073 $\pm$ 0.374)$\times 10^{-1}$ & (8.975 $\pm$ 0.095 $\pm$ 0.634)$\times 10^{-1}$ & {}\\

\hline\hline

\end{tabular}
\label{tab:MeanPtTable}
\end{center}
\end{sidewaystable}

\begin{figure*}[tb!]
  \begin{flushleft}
    \center{\includegraphics[width=0.865\textwidth]{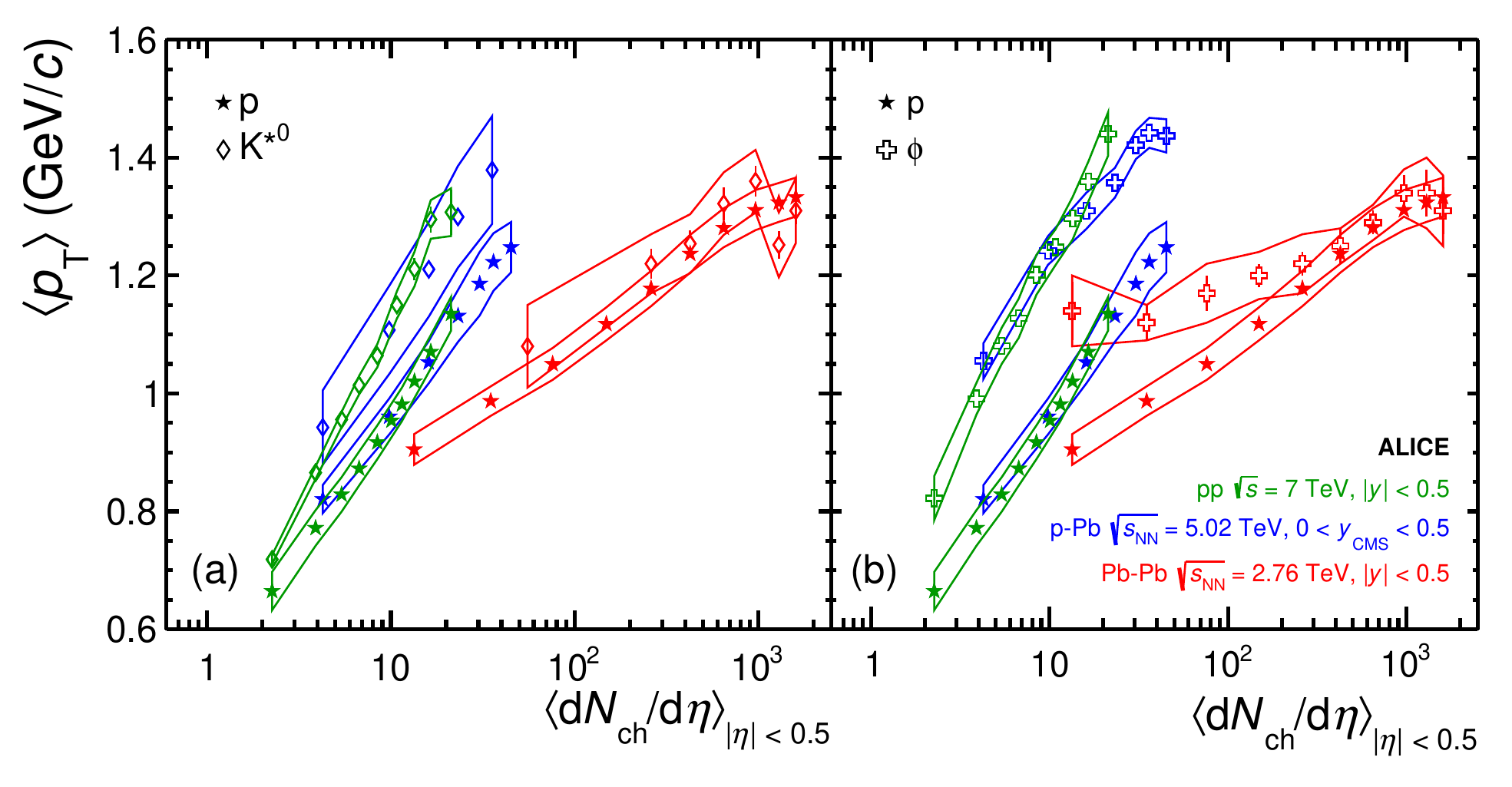}}
    \end{flushleft}
    \caption{Average transverse momentum (a) \kstar\ (left) and (b) $\phi$ compared to protons as a function of 
    charged-particle multiplicity density.} 
    \label{fig:MeanPtResonancesVsProton}
\end{figure*}

The \avpT\ values are observed to increase with multiplicity for all measured particle species, with the increase 
being more pronounced for heavier particles. This observation resembles that of previous measurements in 
\pPb~\cite{Abelev:2013haa} and \PbPb~\cite{Abelev:2013vea} in which \avpT\ values exhibit mass-ordering 
for \pipm, \kapm, \p, \pbar, $\lmb$, $\almb$, \X, \Ix, \Om\ and \Mo. However, while in central \PbPb\ collisions
particles with similar mass such as $\phi$ and \p\ had similar \avpT for central collisions, this is not the case 
for high-multiplicity pp events, where the \avpT\ of the $\phi$  is significantly higher compared to that of the \p, as can be seen in 
Fig.~\ref{fig:MeanPtResonancesVsProton}. This has also been observed in inelastic pp and in p-Pb \cite{Adam:2016bpr} 
and is further indication that radial flow is the dominant mechanism
determining spectra shapes only in very high multiplicity \PbPb. It is also interesting to note that 
at similar multiplicities \avpT\ values for identified particles in pp and \pPb\ are compatible within uncertainties 
despite differences in initial state and collision energy, pointing to a common mechanism at play in these two systems. 

\begin{figure*}[t!]
  \begin{flushleft}
    \center{\includegraphics[width=0.675\textwidth]{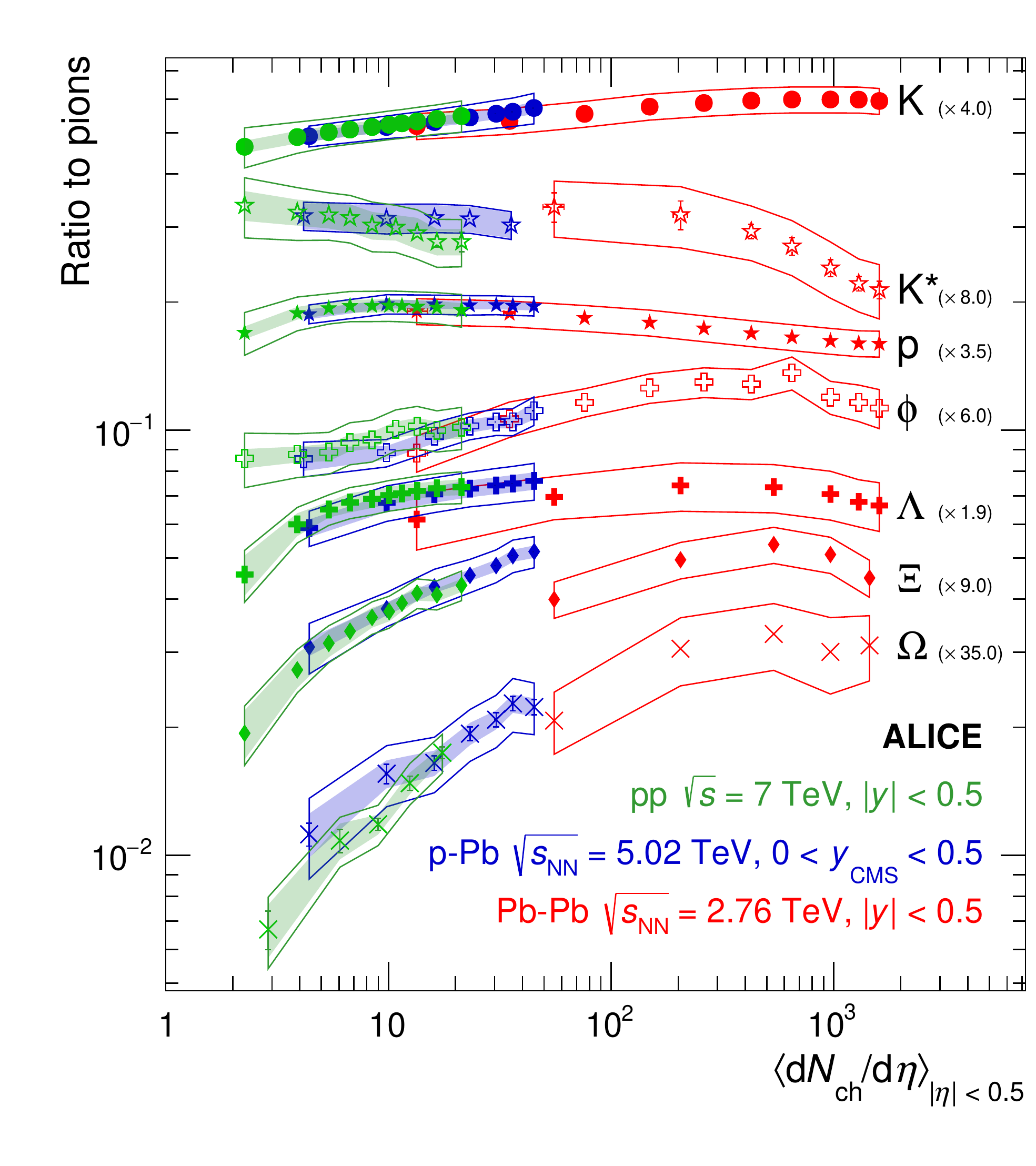}}
    \end{flushleft}
    \caption{Integrated particle-to-pion ratios as a function of \avdNdeta\ for pp, p-Pb and Pb-Pb collisions. Reference p-Pb and Pb-Pb data from \cite{Abelev:2013haa, Adam:2015vsf, 2016720, Abelev:2013xaa,2014196}.}
    \label{fig:dNdyRatios}
\end{figure*}

The multiplicity dependence of identified particle yields relative to pions is compared 
to \pPb\ and \PbPb\ results in Fig.~\ref{fig:dNdyRatios}. Particles with a larger strangeness content are observed
to be produced more abundantly with multiplicity, as can be seen in the $\Lambda/\pi$, $\Xi/\pi$ and $\Omega/\pi$ 
ratios~\cite{Adam:2016emw}. The \p/$\pi$ ratio is observed to be constant within uncertainties
 except for the lowest multiplicity event class. This indicates 
that the increase of hyperon production with respect to pions is a phenomenon that does not originate 
from mass differences but is connected to strangeness content. Furthermore, the relative production of the $\phi$ increases with \dNdeta\ by approximately 20\%, similarly to the
\lmb, a single-strange baryon. This suggests that $\phi$ production cannot be
described solely by considering net strangeness or number of strange quark constituents. The
\kstar/$\pi$ ratio, on the other hand, exhibits a hint of a decrease with multiplicity. In nuclear collisions, 
this decrease is more pronounced and is typically considered a consequence of the rescattering of 
\kstar\ decay daughters during a hadronic phase of the system evolution \cite{Abelev:2014uua,Bleicher:2003ij,Markert:2002rw}. 
All these observations are consistent with previous measurements in \pPb~\cite{Abelev:2013haa,Adam:2015vsf,Adam:2016bpr} and indicate that relative particle abundances can be described as a universal function of charged-particle density in 
pp and \pPb\ collisions. 

\section{\label{secDiscussion}Discussion}

\subsection{\label{subsecMC}Comparison to Monte Carlo models}


\begin{figure*}[t!]
  \begin{flushleft}
    \center{\includegraphics[width=\textwidth]{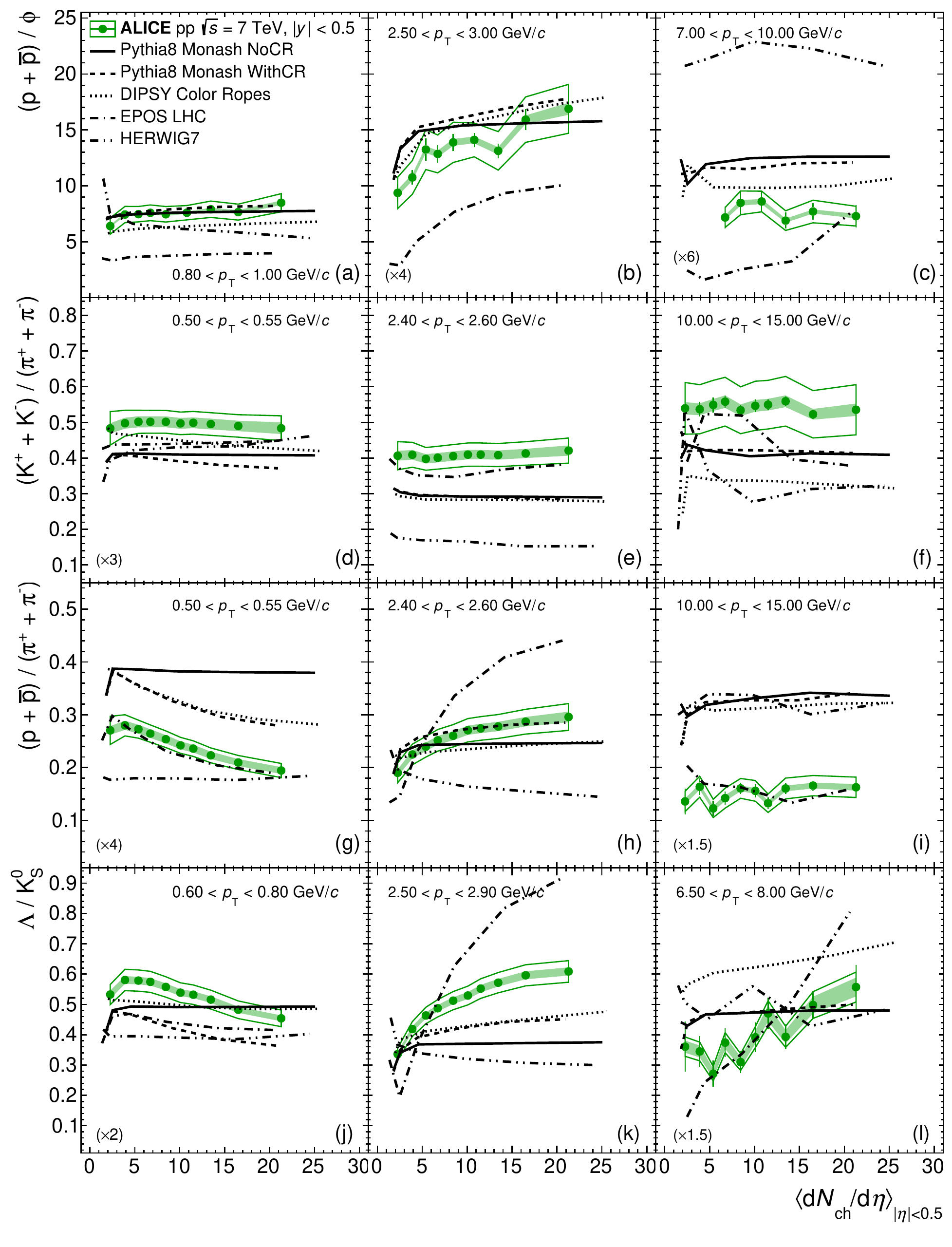}}
    \end{flushleft}
    \caption{Transverse momentum dependence of (a, b, c) \p/$\phi$~=~(p + \pbar)/$\phi$, (d, e, f) \kpi~=~(\kap + \kam)/(\pip + \pim),  (g, h, i) \ppi~=~(p + \pbar)/(\pip +
\pim) and (j, k, l) \lmb/\kzero\ yield ratios in (a, d, g, j) low-, (b, e, h, k) mid- and (c, f, i, l) high-\pt\ intervals compared to several Monte Carlo event generators. 
    Statistical uncertainties are shown as error bars, total systematic uncertainties are shown as hollow bands and multiplicity-uncorrelated 
    systematic uncertainties are shown as shaded bands.}
    \label{fig:RatiosVsPtVsMC}
\end{figure*}

The multiplicity dependence of the \pt-differential \KTopi, \pTopi\ and \LtoKzero\ ratios 
 is compared to several Monte Carlo (MC) event generators, see Fig.~\ref{fig:RatiosVsPtVsMC}. 
 All predictions are obtained by performing selections on charged-particle 
 multiplicities in the V0M acceptance and are compared to data parametrically as a function of \avdNdeta, 
 as discussed in section \ref{subsecEventSelection}. 
 The PYTHIA8 
event generator in its Monash tune~\cite{Sjostrand:2007gs, Skands:2014pea} is only successful in describing the 
qualitative features of the evolution of the
baryon-to-meson ratios if color reconnection (CR) is allowed to occur, as observed already in previous 
work~\cite{PhysRevLett.111.042001}. 
In contrast to the string-model based PYTHIA, the HERWIG code implements hadronization in a clustering approach~\cite{Bahr:2008pv}. As shown in Fig.~\ref{fig:RatiosVsPtVsMC}, the abilities of HERWIG7 to describe particle production at low and intermediate \pt\ are still limited, but are currently being improved~\cite{Gieseke:2016fpz}.
DIPSY, a model in which fragmenting strings are allowed to form 
color ropes which then hadronize with a higher effective string tension~\cite{Bierlich:2014xba}, is also able to 
reproduce the decreasing (increasing) trend of the baryon-to-meson ratios at low (high) \pt, but fails in
describing the absolute values of these ratios. Furthermore, the EPOS-LHC event 
generator, which relies on parton-based Gribov-Regge theory and includes 
elements from hydrodynamics~\cite{Pierog:2013ria}, predicts increased baryon-to-meson ratios at intermediate
\pt\ as a consequence of radial flow, but overestimates the multiplicity dependence of these ratios and thus
fails to quantitatively reproduce the measurements reported here. 
Finally, essentially all models are able to reproduce the fact that the \KTopi\ ratio is multiplicity-independent, 
while not necessarily describing the absolute value well. These findings suggest that there is more than one 
physical mechanism that would lead to the dynamics
observed in \pt-differential identified particle ratios, and a systematic study of other observables
such as the flow coefficients $v_{n}$ is required in order to discriminate among the various possibilities. 

\begin{figure*}[t!]
  \begin{flushleft}
    \center{\includegraphics[width=\textwidth]{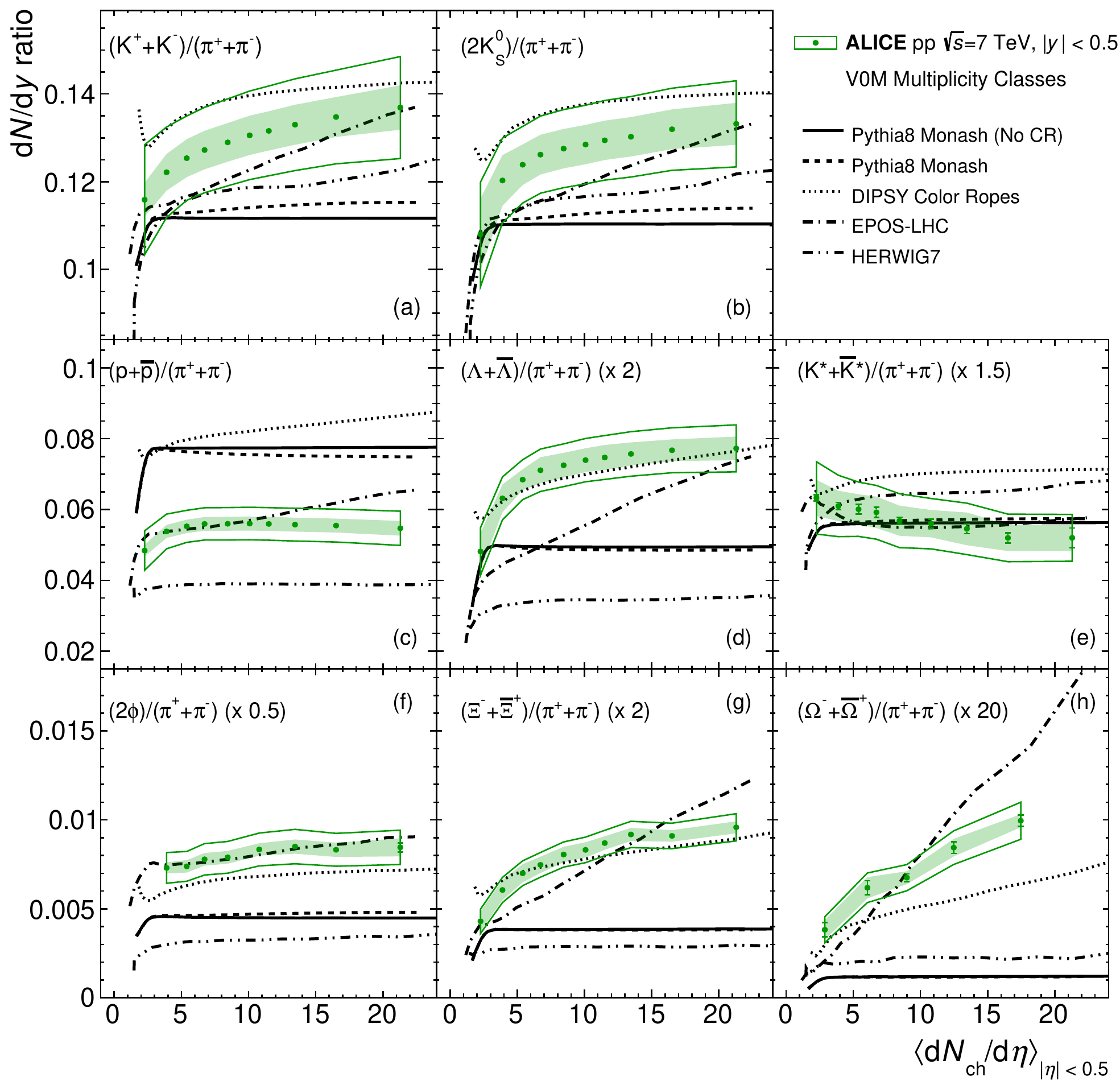}} 
    \end{flushleft}
    \caption{Integrated yield ratios as a function of charged-particle multiplicity density compared to several 
Monte Carlo event generators. Statistical uncertainties are shown as error bars, total systematic uncertainties are shown as hollow bands and multiplicity-uncorrelated 
    systematic uncertainties are shown as shaded bands. For a complete description, please refer to the text.} 
    \label{fig:dNdyRatiosVsMC}
\end{figure*}  

The \pt-integrated ratios are compared to predictions from the same Monte Carlo event generators 
in Fig.~\ref{fig:dNdyRatiosVsMC}. The PYTHIA8 and HERWIG generators incorrectly predict no multiplicity dependence of
relative strangeness production and therefore fail especially in the description of multi-strange baryon production, 
while DIPSY and EPOS-LHC exhibit increased strangeness in high-multiplicity 
pp collisions but fail to predict a flat \pTopi\ ratio. None of the tested generators correctly reproduces
the multiplicity dependence of the \kstar/\pion\ ratio, which is observed to decrease slightly. This comprehensive set of measurements provides essential input for all models aiming to describe flavor generation in \pp\ collisions.

\subsection{\label{subsecBW}Comparison to Blast-Wave model predictions}

\begin{figure*}[t!]
  \begin{flushleft}
    \center{\includegraphics[width=0.605\textwidth]{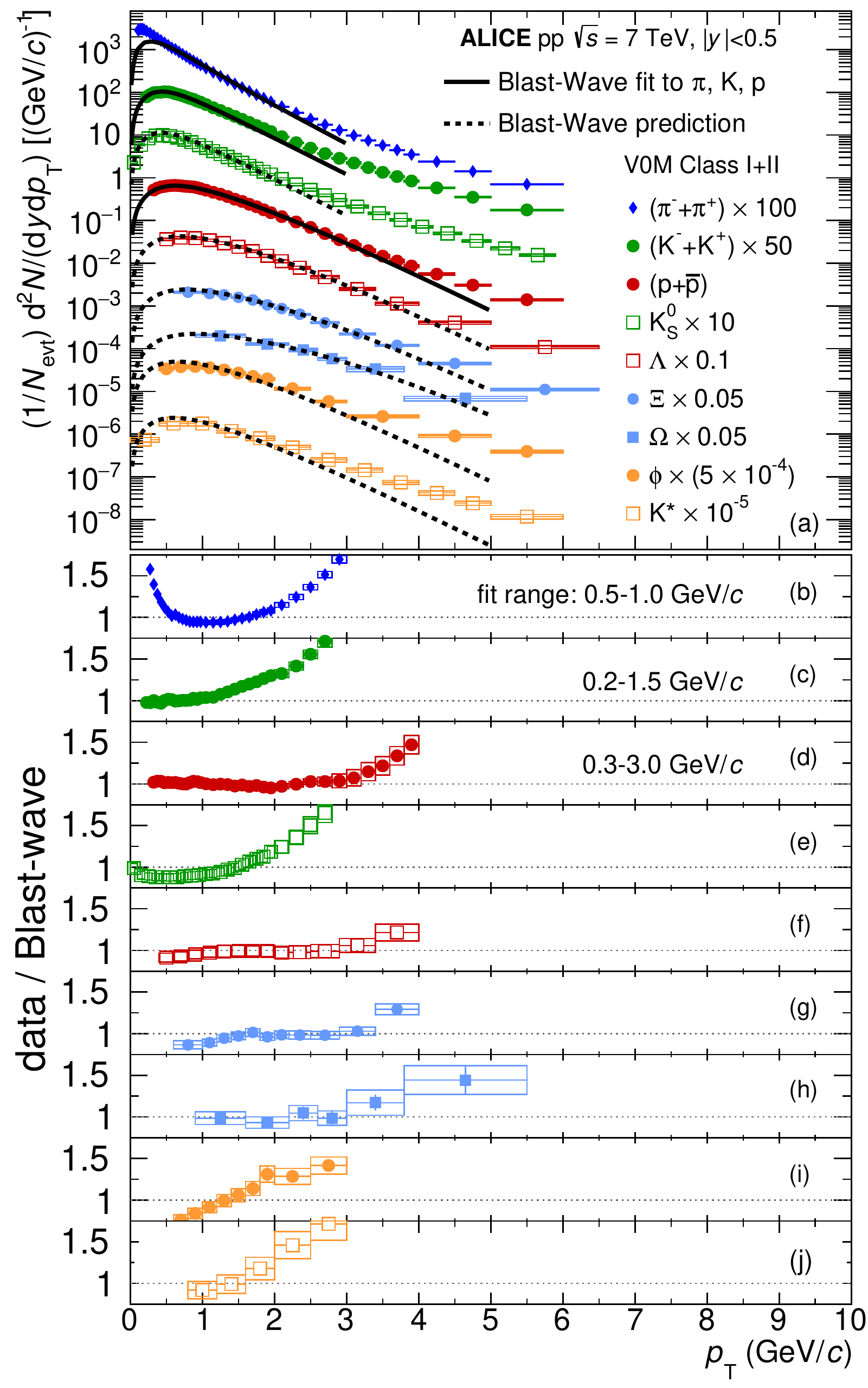}}
    \end{flushleft}
    \caption{(a) Simultaneous BG-BW fit to $\pi$, $K$ and \p\ spectra from high-multiplicity (I+II V0M classes) \pp\ collisions.
             Solid lines correspond to the fit, dashed lines to the prediction. (b, c, d, e, f, g, h, i, j) Ratio data/fit(prediction) for the various
             particle species.}
    \label{fig:SpectraVsBW}
\end{figure*}

The evolution of the \pt\ distribution with multiplicity in \pp\ collisions is remarkably similar to the evolution
observed in larger colliding systems, as underlined in Section~\ref{secResults}.
In larger systems, this evolution can be interpreted as originating from the hydrodynamical radial expansion of the produced 
medium~\cite{Abelev:2013haa,Abelev:2013vea} that can be studied by means of the Boltzmann-Gibbs Blast-Wave model (BG-BW)~\cite{Schnedermann:1993}. 
This model assumes a locally thermalized medium which expands with a
common velocity field and then undergoes an instantaneous freeze-out phase. The average expansion velocity \betaT\ and the kinematic 
freeze-out temperature \Tkin\ can be extracted with a simultaneous fit to the \pt\ distribution of several particles for each multiplicity bin and
the result can be used to predict the \pt\ spectra of particles with different masses.\\
The result of the simultaneous BG-BW fit to $\pi$, $K$ and \p\ for the combined I+II V0M multiplicity class in \pp\ collisions at \seven\ is shown as solid lines in Fig.~\ref{fig:SpectraVsBW} (top) for the \pt\ ranges 0.5-1 GeV/$c$, 0.2-1.5 GeV/$c$ and 
0.3-3 GeV/$c$ respectively.
The predicted spectra for \kzero, \lmb, \sXi, \sOmega, \kstar\ and $\phi$ are shown as
dashed lines. The ratios between the data points and the fits or predictions are plotted in the bottom panels. Strange and multi-strange hadron spectra are reasonably 
well predicted by the BG-BW model in a \pt\ range which gets larger as the mass of the particle increases. This indicates that strange particles may
follow a common motion together with lighter hadrons and suggests the presence of radial flow even in \pp\ collisions. Resonances seem not to follow a similar expansion 
pattern, as there is no \pt\ region where the ratio data/prediction is flat. This discrepancy between BG-BW predictions and resonance 
spectra extends to all multiplicity intervals studied. 

\begin{figure*}[t!]
  \begin{flushleft}
    \center{\includegraphics[width=0.69\textwidth]{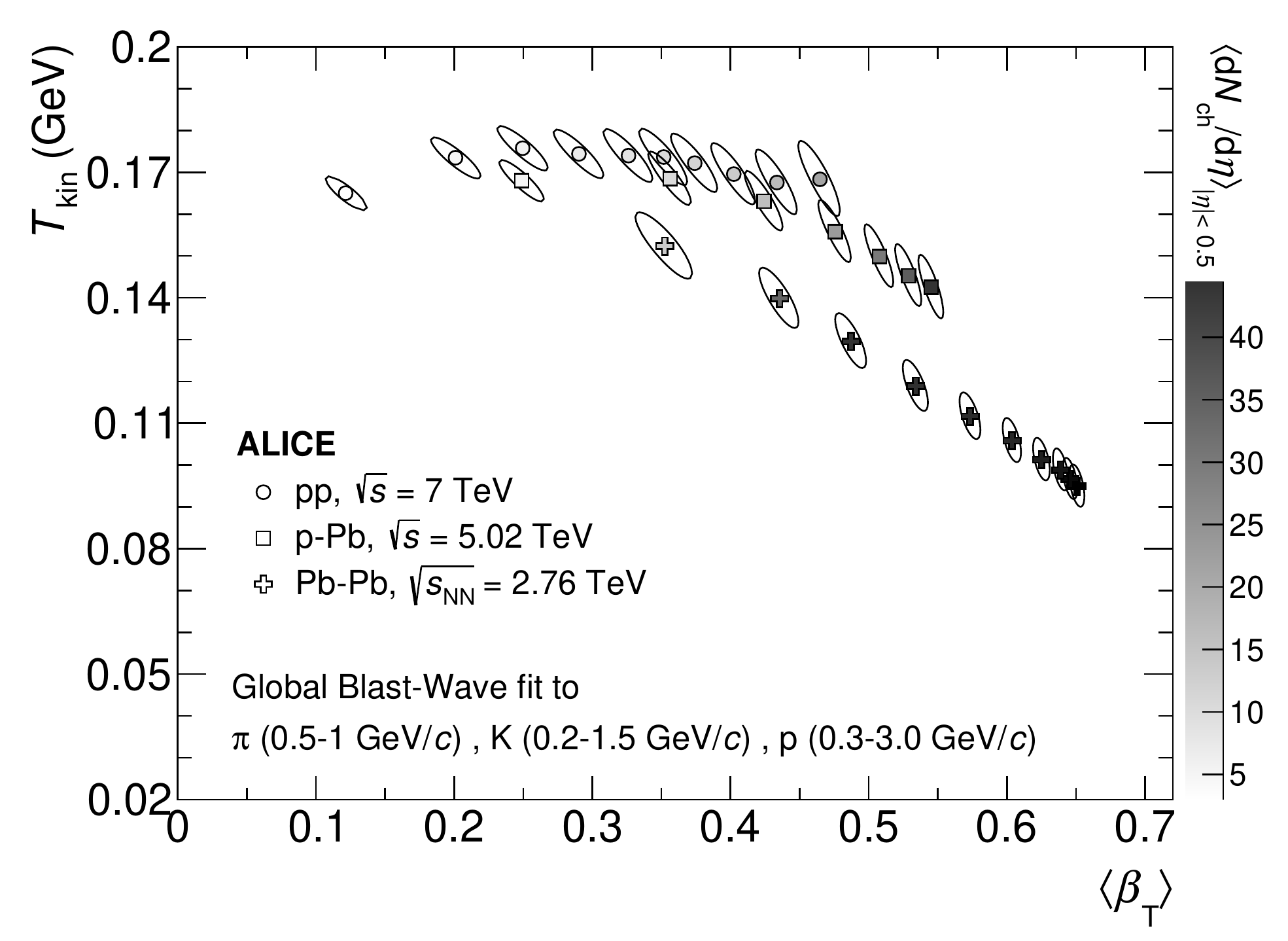}}
    \end{flushleft}
    \caption{Kinematic freeze-out temperature parameter \Tkin\ versus average expansion velocity \betaT\ from a simultaneous BG-BW fit to $\pi$, $K$ and \p\
             spectra measured in \pp, \pPb\ and \PbPb\ collisions. The shade of the datapoints indicates the corresponding average charged-particle 
             multiplicity density.}
    \label{fig:BWbetaT}
\end{figure*}

The fit to \pp\ spectra of $\pi$, $K$ and \p\ has been performed for all the analyzed multiplicity bins and the values of \betaT\
and \Tkin\ are compared to those obtained for \pPb\ and \PbPb\ collisions in Fig.~\ref{fig:BWbetaT}. 
The fitting ranges are the same for all the three systems and a common color palette is used to highlight the average multiplicity 
corresponding to each point. Ellipses correspond to the 1-sigma contour, estimated by fitting the \pt-differential spectra with total uncertainties, i.e. after 
adding statistical and systematic uncertainties in quadrature.

Spectra in \pp\ and \pPb\ lead to very similar \betaT\ and \Tkin\ values when considering similar multiplicities, while in \PbPb\, at similar 
multiplicities, lower \betaT\ are 
observed with respect to the other two systems. This behavior has been interpreted to be a consequence of stronger radial flow gradients in the smaller 
collision systems \cite{PhysRevC.88.044915}. 

The CMS Collaboration has recently reported a similar study in~\cite{CMS:2016str}, where a BG-BW fit to the spectra of \kzero\ and \lmb\ 
has been performed for the three colliding systems, selecting the multiplicity with a central rapidity estimator. 
The \betaT-\Tkin\ progression is found to be different for \pp\ and \pPb\ collisions at high multiplicity, but the numerous differences 
with respect to the analysis discussed here (multiplicity estimator, particles included and treatment of systematic errors in the fits) 
do not allow for a quantitative comparison.

\subsection{\label{subsecSHM}Comparison to statistical hadronization models}

The measured abundances of hadrons produced in heavy-ion collisions have been successfully described by statistical hadronization models over a wide range of energies~\cite{Andronic:2011yq,Cleymans:1998fq,Andronic:2005yp}. Statistical model calculations for central ultra-relativistic heavy-ion collisions are typically carried out in the grand-canonical ensemble. However, a grand-canonical description of particle production is only applicable if the volume $V=R^{3}$ of the system is sufficiently large and as a rule of thumb one needs $VT^{3} > 1$ for a grand-canonical description to hold~\cite{Hagedorn:1984uy,Rafelski:1980gk}. This condition must be fulfilled for each conserved charge separately.
Several attempts were made to extend the picture of statistical hadronization to smaller systems such as pp or even e$^{+}$e$^{-}$ collisions~\cite{Redlich:2009xx,Becattini:1996gy,Kraus:2008fh}. While particle ratios of non-strange particles are observed to be similar in small and large systems, the relative production of strange particles appears to be significantly lower in smaller systems. The data presented in~\cite{Adam:2016emw} show for the first time that there is a continuous increase of strangeness production with increasing event multiplicity across various collision systems. In the strangeness canonical approach, it is assumed that the total amount of strange hadrons in the volume is small with respect to non-strange hadrons. Thus the conservation of strangeness is guaranteed locally and not only globally while the bulk of the particles is still described in the grand-canonical ensemble. For further details on this approach, we refer for instance to~\cite{BraunMunzinger:2003zd,Hamieh:2000tk,Cleymans:2004bf}. The study presented here utilizes the implementation in the THERMUS 3.0 code~\cite{Wheaton:2004qb}. Alternative implementations of the statistical model are for instance adopted in the codes of the GSI-Heidelberg~\cite{Andronic:2005yp,Andronic:2017pug} and SHARE~\cite{PETRAN20142056} groups.


\subsubsection{Correlation volume for strangeness production} 

Previous studies based on THERMUS were targeted to describe the evolution of multi-strange particle production in p-Pb collisions as a function of event multiplicity~\cite{Adam:2015vsf}. In this case, the volume for particle production was chosen such that the charged pion multiplicity d$N_{\pi}$/d$y$ at midrapidity corresponding to a window of $y = \pm 0.5$ units was correctly described by the model. This approach is equivalent to calculating strangeness suppression for a  system whose total extension only corresponds to one unit of rapidity. Consequently, the model describes qualitatively the suppression pattern, but overestimates the reduction of strangeness at small multiplicities. The discrepancy increases even further if this approach is extended to the smaller multiplicities which are covered in pp collisions.

\begin{figure*}[t!]
  \begin{flushleft}
    \center{\includegraphics[width=\textwidth]{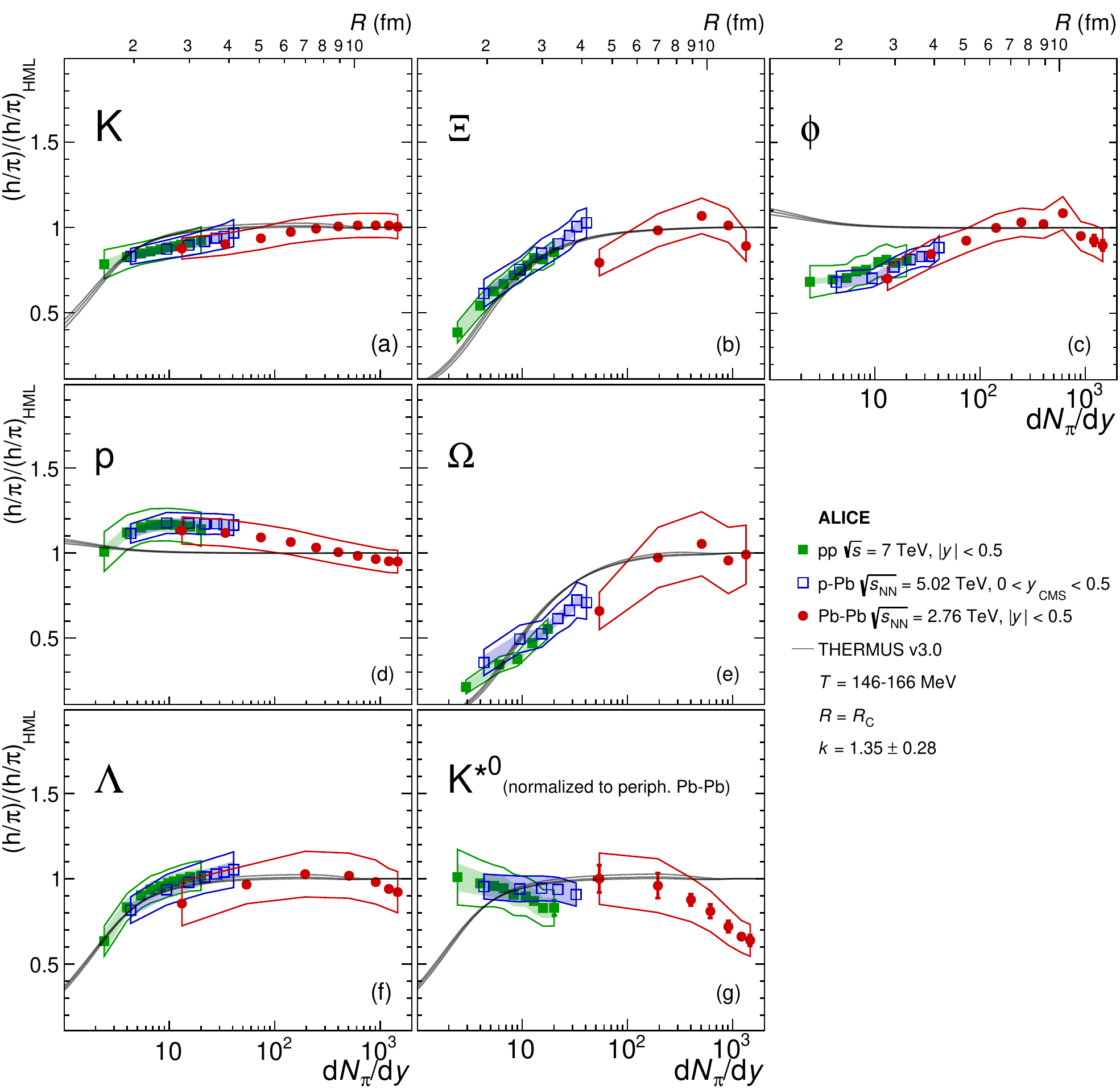}}
    \end{flushleft}
    \caption{Ratios of the production yields to pions for several particle species as a function of the midrapidity pion yields for pp, \pPb\ and \PbPb\ colliding systems compared to the THERMUS model prediction for the strangeness canonical suppression (continuous curves), in which only the system size is varied.  All values except for the \kstar\ are normalized to the high multiplicity limit (see text for details). The upper axis shows the radius $R$ of the correlation volume $V=R^{3}$ which corresponds to the predicted particle ratios. The variations in model prediction curves correspond to variations of the chemical freeze-out temperature between 146, 156 and 166~MeV. Reference p-Pb and Pb-Pb data from \cite{Abelev:2013haa, Adam:2015vsf, 2016720, Abelev:2013xaa,2014196}.
    }
    \label{fig:SHM_Suppression}
\end{figure*}

In the study presented here, a similar approach as in~\cite{Adam:2015vsf} is followed: the strangeness saturation parameter is fixed to $\gamma_S = 1$, the chemical potentials associated to baryon and electric charge quantum numbers are set to zero. The chemical freeze-out temperature $T_{ch}$ is varied from 146 to 166~MeV as in~\cite{Adam:2015vsf} following recent results from  Lattice QCD calculations and their uncertainties~\cite{Bazavov:2014pvz} as well as by fits to experimental data from central Pb-Pb collisions~\cite{Acharya:2017bso,Andronic:2017pug}. Ratios of the production yields to pions are investigated for several particle species. In order to cancel the influence of the freeze-out temperature and to isolate the volume dependence, all ratios except for \kstar\ are normalized to the high multiplicity limit, i.e. the grand-canonical saturation value for the model and the mean ratio in the 0-60\% most central Pb-Pb collisions for the data.
As the production rates of short-lived resonances in central heavy-ion collisions might be reduced by re-scattering effects in the hadronic phase~\cite{Abelev:2014uua,Adam:2016bpr}, the values for K$^{0*}$ were normalized to the most peripheral bin in Pb-Pb collisions. The resulting double ratios are shown in Fig.~\ref{fig:SHM_Suppression}.

In contrast to~\cite{Adam:2015vsf}, the total volume $V$ of the fireball was determined differently. While the strangeness conservation volume is also imposed to be of the size of the fireball ($V=V_C$), its absolute magnitude is not fixed to reproduce the pion multiplicity in a window of $y = \pm 0.5$ at midrapidity. Instead, it is fixed to reproduce a pion multiplicity which is larger by a factor $k$ and thus corresponds to a larger rapidity window assuming a flat dependence of particle production as a function of rapidity. The same factor $k$ was used for all particles and multiplicities. In practice, $k$ corresponds to a constant scaling factor of d$N_{\pi}$/d$y$ (the x-axis in Fig.~\ref{fig:SHM_Suppression}), and this feature is used for its numerical determination: the exact value of $k$ was optimized in a one-parameter fit of the functions describing the evolution of the double ratios versus d$N_{\pi}$/d$y$ to the experimental data.  For the determination of the systematic uncertainty on the value of $k$, an alternative normalization scheme for the data was applied (normalization to the highest available centrality bin) and the procedure for the determination of $k$ was repeated. A value of $k=1.35 \pm 0.28$ is obtained corresponding to a rapidity window of $y = \pm 0.67$. The results thus indicate that the total correlation volume for strangeness production extends over about 1.35 units in rapidity. In other words, strangeness as a conserved quantity in QCD can be effectively equilibrated over this distance in the system. Similar values can be obtained from purely theoretical considerations on causality constraints~\cite{Castorina:2013mba}. Furthermore, the size of the correlation window is also compatible with similar estimates for charm production~\cite{Andronic:2003zv,Andronic:2006ky}. We also note that this value is smaller than the plateau in the rapidity distribution at LHC energies which extends typically over three to four units~\cite{Adam:2015gka,Abbas:2013bpa} and is thus meaningful from a physical point of view.

\subsubsection{Comparison to experimental data} 

As shown in Fig.~\ref{fig:SHM_Suppression} this approach allows for a qualitative description of particle ratios as a function of event multiplicity. They can be naturally described within the framework of strangeness canonical suppression. The deviations observed for the \kstar\ meson in central Pb-Pb collisions can be ascribed to the aforementioned re-scattering effects in the hadronic phase~\cite{Knospe:2015nva}. Furthermore, differences between the model and the data in the most peripheral Pb-Pb collisions in case of $\Omega$ and $\Xi$ can be potentially reduced with core-corona corrections~\cite{Aichelin:2008mi}.

From a quantitative point of view, essentially all data points can be described within 1-2 standard deviations. A potentially different trend is only observed for the $\phi$-meson for which -- as a strangeness-neutral particle -- a flat dependence as a function of event multiplicity is expected from the model, but which shows a rising and falling trend in data. Future experimental data will be needed in order to clarify the significance of this deviation. It must be noted, though, that the $\phi$-meson also deviates from a common blast-wave fit to other light-flavor hadrons in peripheral Pb-Pb collisions indicating an out-of-equilibrium production except for most central Pb-Pb collisions~\cite{Abelev:2014uua}.

Independently of the experimental precision and possible higher order effects in the particle production mechanisms, we find that the strangeness canonical approach can reproduce the multiplicity dependence of all measured 
light-flavor hadrons across various collision systems to within 10-20\%.

\section{\label{secConclusion} Summary and conclusions}

A comprehensive set of unidentified and identified \pt-differential particle spectra at mid-rapidity has been measured in \pp\ collisions
 at \seven\ as a function of charged-particle multiplicity, complementing the existing measurements
in \pPb\ and \PbPb\ collisions and allowing for a comparison of these different collision systems. 
In \pp\ collisions, all transverse momentum spectra are observed to become harder with progressively 
larger charged-particle multiplicity density, with the effect being more pronounced for heavier particles. 
In addition, the modification of spectra with respect to the inclusive measurement follows a different
pattern for mesons and baryons, except for resonances, which follow baryons at a low-\pt\ of up to approximately 2~\gevc and tend to 
be modified similarly to mesons above a \pt\ of 2~\gevc. Furthermore, it has been 
demonstrated that the evolution of the baryon/meson ratios as a function of \avdNdeta\ exhibits
a universal pattern for all collision systems. This behavior might indicate a common mechanism at work that 
depends solely on final-state multiplicity density. A similar statement can 
also be made for integrated particle ratios, which are observed to depend on \avdNdeta\ in approximately the
same way for any colliding system despite crucial differences in the initial states as well as colliding 
energies. The simplest interpretation of this similarity is that in both cases the final state is a thermalized 
system whose volume at hadronization is proportional to the charged-particle multiplicity density. 

In order to test the assertion of equilibration more quantitatively, the Blast-Wave model was employed to check if the 
hypothesis of kinematic equilibrium can describe \pt-differential particle spectra at low momenta, and a statistical hadronization model 
employing a strangeness-canonical approach was used to check for chemical equilibrium. In all cases, all particle 
species except for resonances are 
described within 10-20\% by these models. It is also interesting to note that, within the
statistical hadronization model employed here, the correlation volume over which 
strangeness production is equilibrated extends over approximately 1.35 units of rapidity.

The results are also compared to predictions from event generators, which are only able to describe the
evolution of \pt-differential particle spectra with \avdNdeta\ if mechanisms such as color reconnection, color ropes or radial flow, as is the
case for PYTHIA, DIPSY and EPOS-LHC, respectively, were present. 
To distinguish between these different mechanisms requires more studies, both experimental and theoretical.


The multiplicity dependence of 
relative abundances of identified particles was also compared to several 
event generators and it was found that no generator is able to fully describe the whole observed dynamics satisfactorily. This comprehensive set of results therefore provides a challenge to 
the theory community and
represents an opportunity to study not only \pp\ collisions specifically, but also hadronization at high energies in general. 


\newenvironment{acknowledgement}{\relax}{\relax}
\begin{acknowledgement}
\section*{Acknowledgements}

The ALICE Collaboration would like to thank all its engineers and technicians for their invaluable contributions to the construction of the experiment and the CERN accelerator teams for the outstanding performance of the LHC complex.
The ALICE Collaboration gratefully acknowledges the resources and support provided by all Grid centres and the Worldwide LHC Computing Grid (WLCG) collaboration.
The ALICE Collaboration acknowledges the following funding agencies for their support in building and running the ALICE detector:
A. I. Alikhanyan National Science Laboratory (Yerevan Physics Institute) Foundation (ANSL), State Committee of Science and World Federation of Scientists (WFS), Armenia;
Austrian Academy of Sciences and Nationalstiftung f\"{u}r Forschung, Technologie und Entwicklung, Austria;
Ministry of Communications and High Technologies, National Nuclear Research Center, Azerbaijan;
Conselho Nacional de Desenvolvimento Cient\'{\i}fico e Tecnol\'{o}gico (CNPq), Universidade Federal do Rio Grande do Sul (UFRGS), Financiadora de Estudos e Projetos (Finep) and Funda\c{c}\~{a}o de Amparo \`{a} Pesquisa do Estado de S\~{a}o Paulo (FAPESP), Brazil;
Ministry of Science \& Technology of China (MSTC), National Natural Science Foundation of China (NSFC) and Ministry of Education of China (MOEC) , China;
Ministry of Science and Education, Croatia;
Ministry of Education, Youth and Sports of the Czech Republic, Czech Republic;
The Danish Council for Independent Research | Natural Sciences, the Carlsberg Foundation and Danish National Research Foundation (DNRF), Denmark;
Helsinki Institute of Physics (HIP), Finland;
Commissariat \`{a} l'Energie Atomique (CEA) and Institut National de Physique Nucl\'{e}aire et de Physique des Particules (IN2P3) and Centre National de la Recherche Scientifique (CNRS), France;
Bundesministerium f\"{u}r Bildung, Wissenschaft, Forschung und Technologie (BMBF) and GSI Helmholtzzentrum f\"{u}r Schwerionenforschung GmbH, Germany;
General Secretariat for Research and Technology, Ministry of Education, Research and Religions, Greece;
National Research, Development and Innovation Office, Hungary;
Department of Atomic Energy Government of India (DAE), Department of Science and Technology, Government of India (DST), University Grants Commission, Government of India (UGC) and Council of Scientific and Industrial Research (CSIR), India;
Indonesian Institute of Science, Indonesia;
Centro Fermi - Museo Storico della Fisica e Centro Studi e Ricerche Enrico Fermi and Istituto Nazionale di Fisica Nucleare (INFN), Italy;
Institute for Innovative Science and Technology , Nagasaki Institute of Applied Science (IIST), Japan Society for the Promotion of Science (JSPS) KAKENHI and Japanese Ministry of Education, Culture, Sports, Science and Technology (MEXT), Japan;
Consejo Nacional de Ciencia (CONACYT) y Tecnolog\'{i}a, through Fondo de Cooperaci\'{o}n Internacional en Ciencia y Tecnolog\'{i}a (FONCICYT) and Direcci\'{o}n General de Asuntos del Personal Academico (DGAPA), Mexico;
Nederlandse Organisatie voor Wetenschappelijk Onderzoek (NWO), Netherlands;
The Research Council of Norway, Norway;
Commission on Science and Technology for Sustainable Development in the South (COMSATS), Pakistan;
Pontificia Universidad Cat\'{o}lica del Per\'{u}, Peru;
Ministry of Science and Higher Education and National Science Centre, Poland;
Korea Institute of Science and Technology Information and National Research Foundation of Korea (NRF), Republic of Korea;
Ministry of Education and Scientific Research, Institute of Atomic Physics and Romanian National Agency for Science, Technology and Innovation, Romania;
Joint Institute for Nuclear Research (JINR), Ministry of Education and Science of the Russian Federation and National Research Centre Kurchatov Institute, Russia;
Ministry of Education, Science, Research and Sport of the Slovak Republic, Slovakia;
National Research Foundation of South Africa, South Africa;
Centro de Aplicaciones Tecnol\'{o}gicas y Desarrollo Nuclear (CEADEN), Cubaenerg\'{\i}a, Cuba and Centro de Investigaciones Energ\'{e}ticas, Medioambientales y Tecnol\'{o}gicas (CIEMAT), Spain;
Swedish Research Council (VR) and Knut \& Alice Wallenberg Foundation (KAW), Sweden;
European Organization for Nuclear Research, Switzerland;
National Science and Technology Development Agency (NSDTA), Suranaree University of Technology (SUT) and Office of the Higher Education Commission under NRU project of Thailand, Thailand;
Turkish Atomic Energy Agency (TAEK), Turkey;
National Academy of  Sciences of Ukraine, Ukraine;
Science and Technology Facilities Council (STFC), United Kingdom;
National Science Foundation of the United States of America (NSF) and United States Department of Energy, Office of Nuclear Physics (DOE NP), United States of America.
\end{acknowledgement}

\bibliographystyle{utphys} 	
\bibliography{PidSpectraVsMultLong}

\newpage
\appendix

\newpage
\section{The ALICE Collaboration}
\label{app:collab}

\begingroup
\small
\begin{flushleft}
S.~Acharya$^{\rm 139}$, 
F.T.-.~Acosta$^{\rm 20}$, 
D.~Adamov\'{a}$^{\rm 93}$, 
A.~Adler$^{\rm 74}$, 
J.~Adolfsson$^{\rm 80}$, 
M.M.~Aggarwal$^{\rm 98}$, 
G.~Aglieri Rinella$^{\rm 34}$, 
M.~Agnello$^{\rm 31}$, 
N.~Agrawal$^{\rm 48}$, 
Z.~Ahammed$^{\rm 139}$, 
S.U.~Ahn$^{\rm 76}$
% ...

\bigskip

\bigskip 

\textbf{\Large Affiliation Notes}

\bigskip 

% ...
$^{\rm V}$ Also at: Institute of Theoretical Physics, University of Wroclaw, Poland\\

\bigskip

\bigskip 

\textbf{\Large Collaboration Institutes}

\bigskip 

% ...
$^{20}$ Department of Physics, University of California, Berkeley, California, United States\\
% ...
$^{31}$ Dipartimento DISAT del Politecnico and Sezione INFN, Turin, Italy\\
% ...
$^{34}$ European Organization for Nuclear Research (CERN), Geneva, Switzerland\\
% ...
$^{48}$ Indian Institute of Technology Bombay (IIT), Mumbai, India\\
% ...
$^{74}$ Johann-Wolfgang-Goethe Universit\"{a}t Frankfurt Institut f\"{u}r Informatik, Fachbereich Informatik und Mathematik, Frankfurt, Germany\\
% ...
$^{76}$ Korea Institute of Science and Technology Information, Daejeon, Republic of Korea\\
% ...
$^{80}$ Lund University Department of Physics, Division of Particle Physics, Lund, Sweden\\
% ...
$^{93}$ Nuclear Physics Institute of the Czech Academy of Sciences, \v{R}e\v{z} u Prahy, Czech Republic\\
% ...
$^{98}$ Physics Department, Panjab University, Chandigarh, India\\
% ...
$^{139}$ Variable Energy Cyclotron Centre, Homi Bhabha National Institute, Kolkata, India\\
% ...

\bigskip 

\end{flushleft} 

\end{document}